\documentclass[preprint]{aastex}
\usepackage{psfig}

\newcommand{\myemail}{ettoref@oapa.astropa.unipa.it}
\slugcomment{To appear in ApJ}
\shorttitle{The Orion Nebula Cluster. Paper II}
\shortauthors{Flaccomio et al.}

\begin{document}
\title{{\em Chandra} X-ray Observation of the Orion Nebula Cluster. II
Relationship between X-ray activity indicators and stellar parameters}    
\author{E. Flaccomio\altaffilmark{1}}
\affil{Dipartimento di Scienze Fisiche ed Astronomiche --  Universit\`a di 
Palermo}
\email{\myemail}

\author{F. Damiani, G. Micela,  S. Sciortino}
\affil{Osservatorio Astronomico di Palermo G.S. Vaiana, Palazzo
dei Normanni, I-90134 Palermo, Italy}

\and

\author{F. R. Harnden Jr., S. S. Murray , S. J. Wolk} 
\affil{Harvard-Smithsonian Center for Astrophysics, 60 Garden Street, 
Cambridge, MA 02138} 

\altaffiltext{1}{Now at Osservatorio Astronomico di Palermo G.S. Vaiana}


\begin{abstract} 

Using the results of our first paper on the {\em Chandra} HRC observation of
the Orion Nebula Cluster (ONC), here we explore the relation between the
coronal activity of its 1-Myr-old pre-main sequence population and stellar
parameters.  We find that median X-ray luminosities of low mass stars
($M/M_{\odot} \lesssim 3$) increase with increasing mass and decreasing stellar
age. Brown dwarfs ($0.03 \lesssim M/M_{\odot} \lesssim 0.08$) follow the same
trend with mass.  From $M \sim 0.1$ to $M \sim 0.5~M_{\odot}$, median $L_X/L_{bol}$ values increase by about half an order
of magnitude and then remain constant at
$\sim 10^{-3.5}$ for the mass range from 0.5 to 3.0 $M/M_{\odot}$. In these same two
mass ranges, $L_X/L_{bol}$ remains roughly constant with age, until it
drops by more than two orders of magnitudes at the epoch when $\sim 2-4
M_\odot$ stars are expected to become fully radiative.

We find a dependence of $L_X$ and $L_X/L_{bol}$ on circumstellar accretion
indicators and suggest three possible hypotheses for its origin. In spite of
improved X-ray and rotational data, correlations between activity indicators
and rotation remain elusive for these stars, possibly indicating that stars for
which rotational periods have been measured have reached some saturation level.

Our study of X-ray activity vs. stellar mass leads us to propose that
the few HRC X-ray sources not associated with any optical/infrared
counterpart trace a yet to be discovered stellar population of deeply
embedded, relatively massive ONC members.

\end{abstract} 


\section{Introduction}

The present work focuses on the study of the X-ray activity of Orion
Nebula Cluster (ONC) members: from a purely observational standpoint,
we explore the relationship between X-ray activity and stellar
characteristics, searching for the physical mechanisms responsible for
coronal X-ray activity in pre-main sequence (PMS) stars. In spite of
the previous observational work in this area
\citep[e.g.,][]{fei93,cas95,gag95a,fla00a}, no definitive picture has
yet emerged.

Our investigation is based on the original {\em Chandra} High Resolution Camera
\citep[HRC, ][]{mur00} X-ray data and literature optical data  presented in
\citet[][ hereafter Paper~I]{papI} and \citet{fla02} for objects in the ONC
area. For the ONC region under study, Paper~I defined an {\em optical sample}
of 696 optically selected, extinction limited, and well characterized ONC
members. Here we address the controversial questions of the relationship of
X-ray activity with convection and rotation (i.e., the classical
$\alpha-\omega$ dynamo parameters), with stellar mass and age, and with
circumstellar accretion and/or the presence of circumstellar disks. Taking
advantage of our detailed description of activity as a function of stellar
mass, we then explore the nature of the few detected X-ray sources not
associated with any optical/infrared (IR) counterparts.

The structure of this paper is as follows: In \S \ref{sect:XvsOpt} we
study the relation of activity with rotation, mass, age and
circumstellar accretion strength. In \S \ref{sect:unid} we
speculate on the nature of X-ray sources with no optical/IR
counterpart. Finally we discuss our findings and summarize our
conclusions in \S \ref{sect:disc}. An Appendix details our technique for
investigating possible sources below our X-ray sensitivity limit.

\section{Relationships between X-ray activity indicators and stellar parameters
\label{sect:XvsOpt}}

In exploring relationships between X-ray activity and stellar characteristics
through empirical correlations with physically meaningful stellar parameters,
our main goal is to understand the $\sim$5 orders of magnitude spread that we observe in $L_X$ and
$L_X/L_{bol}$.  We focus on how
activity is related to stellar rotational period, to mass and evolutionary
status (as inferred from evolutionary tracks - see Paper~I), and to the
presence of an accretion disk.  Paper~I presents relevant X-ray and optical
data for our {\em optical sample} comprised of 696 ONC likely members, 342 of
which are unambiguously detected in our HRC X-ray data and have both low
optical extinction ($A_V < 3.0$) and available mass estimates.
This stellar sample was carefully selected in order to avoid X-ray
selection effects.

Our non-parametric approach is to split our reference
sample into several bins of the stellar parameter under examination, to
evaluate for each of these bins a Maximum Likelihood distribution function
(using the {\em Kaplan-Mayer} estimator) for our two X-ray activity indicators,
and then to examine the trends of these distributions as a function of the
parameter.
The Kaplan-Mayer estimators were computed using the ASURV package \citep{fei85},
which was also used to perform various two population tests\footnote{Gehan's
generalized Wilcoxon test using both permutation and hypergeometric variance;
Logrank test; Peto and Peto generalized Wilcoxon test; Peto and Prentice
generalized Wilcoxon test.} on pairs of selected stellar subsamples in order to
establish the confidence with which we can reject the null hypothesis (that two
distributions of a given parameter are drawn from the same parent distribution).

\subsection{Rotational Period \label{sect:prot}}

We first searched for a correlation between X-ray activity and stellar
rotational period ($P_{rot}$). Such a correlation is known to hold for older,
main sequence stars of low mass and moderate rotation rate and is the main
evidence for activity being driven by an $\alpha - \omega$ stellar dynamo. Such
correlations have not previously been found for the ONC stars \citep{gag95a}
and several other PMS associations \citep[e.g., ][]{wal91,fei93,alc97}. In the
young Taurus region, to the contrary, a correlation between activity and
rotation has been observed \citep{bou90,str94,dam95,stel01}. Given the
incompleteness of rotational databases used in most previous studies, the
question remains open.

\begin{figure}[!t!]
\centerline{\psfig{figure=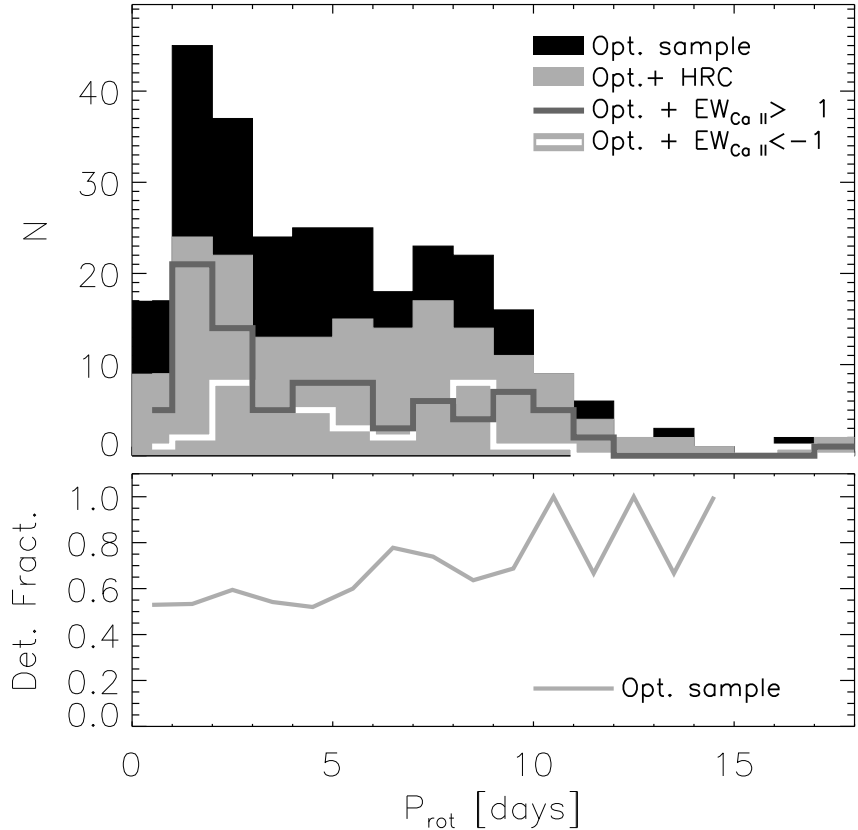,width=12cm}}
\caption{Upper panel: Rotational period distribution for the {\em optical sample}, the subset of stars detected in our HRC data and two subsets of the {\em optical  sample} segregated by their Ca II line equivalent width (see \S \ref{sect:CaII}). Lower panel: detection fraction as a function of $P_{rot}$. \label{fig:P_dist}}
\end{figure}

\begin{figure}[t]
\centerline{\psfig{figure=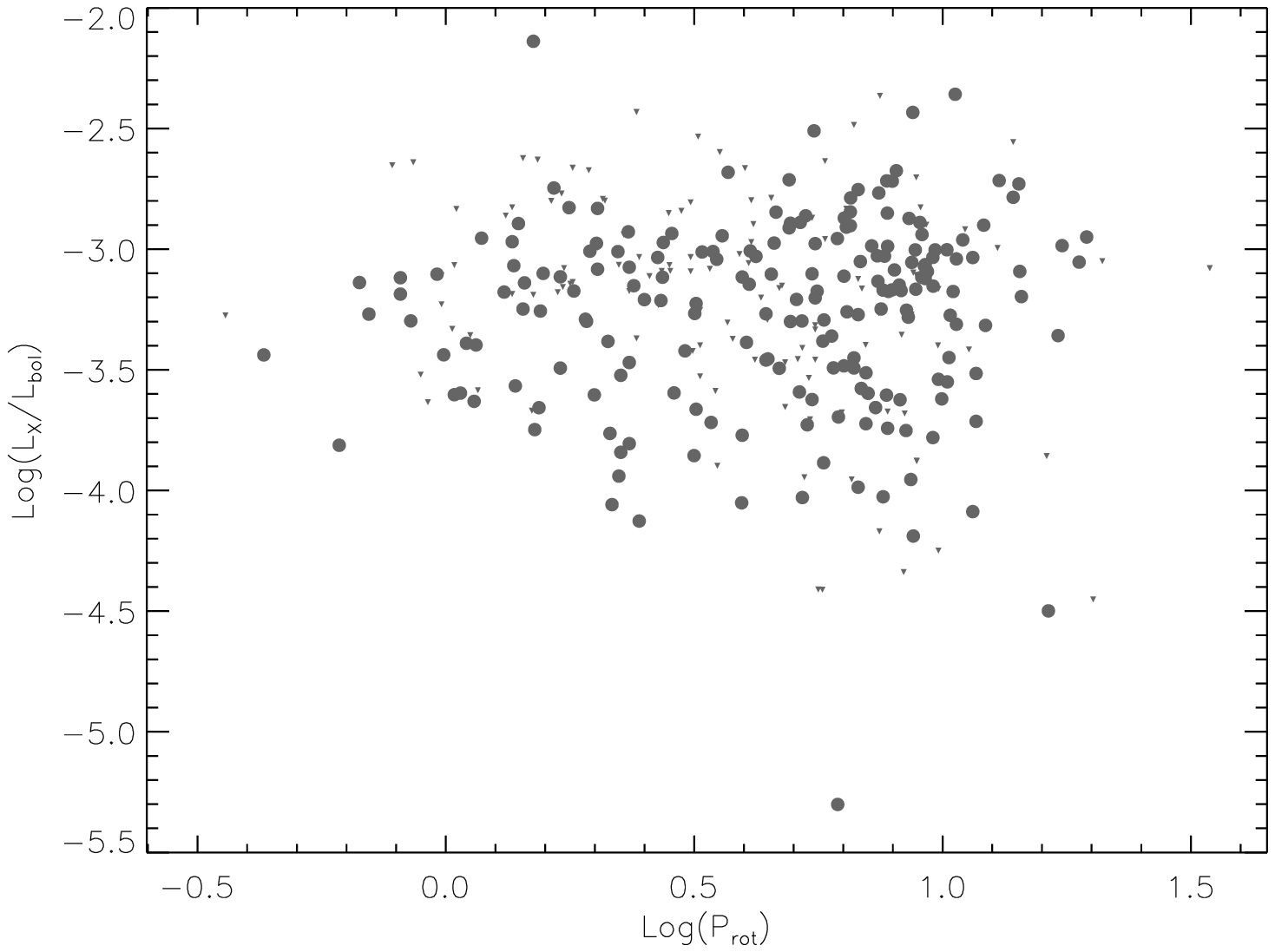,width=15cm}}
\caption{$Log(L_X/L_{bol})$ vs. $Log(P_{rot})$ scatter plot. Filled circles
represent detections, small triangles upper limits in $Log(L_X/L_{bol})$.
\label{fig:LXLbvsProt}}
\end{figure}

In Figure \ref{fig:P_dist} we show the distribution of rotational periods
(upper panel) collected in Paper~I from the recent literature for 282 stars in
our {\em optical sample}. Also shown vs. $P_{rot}$ is the HRC detection
fraction (lower panel). Our database of rotational periods is significantly
more complete than that available to \citet{gag95a}, but in spite of this
improvement and {\em Chandra's} higher X-ray sensitivity (which yields a high
detection fraction), we still can find no evidence of correlation between X-ray
luminosity and $P_{rot}$. The scatter plot of $Log(L_X/L_{bol})$ vs.
$Log(P_{rot})$ (Figure \ref{fig:LXLbvsProt}) similarly shows no evidence of a
correlation with $P_{rot}$. These results hold true no matter what physically
meaningful subsample of the whole we consider.  Indeed, in order to minimize
stellar structure differences throughout the sample that might wash out a
correlation with $P_{rot}$, we repeated the analysis in several mass bins (see
\S \ref{sect:LxvsMass}) and also for high and low accretion stars (see \S
\ref{sect:CaII}), obtaining the same negative results in all cases.

It should be noted that the sample of stars for which rotational periods are
available is strongly biased toward the most X-ray active stars. For example,
the median $Log(L_X/L_{bol})$ for stars with known rotational period and mass
smaller than 3.0$M_{\odot}$ is -3.37 while the corresponding median for stars
of the same mass with unknown rotational periods is -3.73.  Moreover, the
subsample of stars with known $P_{rot}$ appears to be dominated by the most
weakly accreting stars in the whole sample (see \S \ref{sect:CaII}).



\subsection{Stellar mass stars \label{sect:LxvsMass}}

\begin{figure}[!h!]
\centerline{\psfig{figure=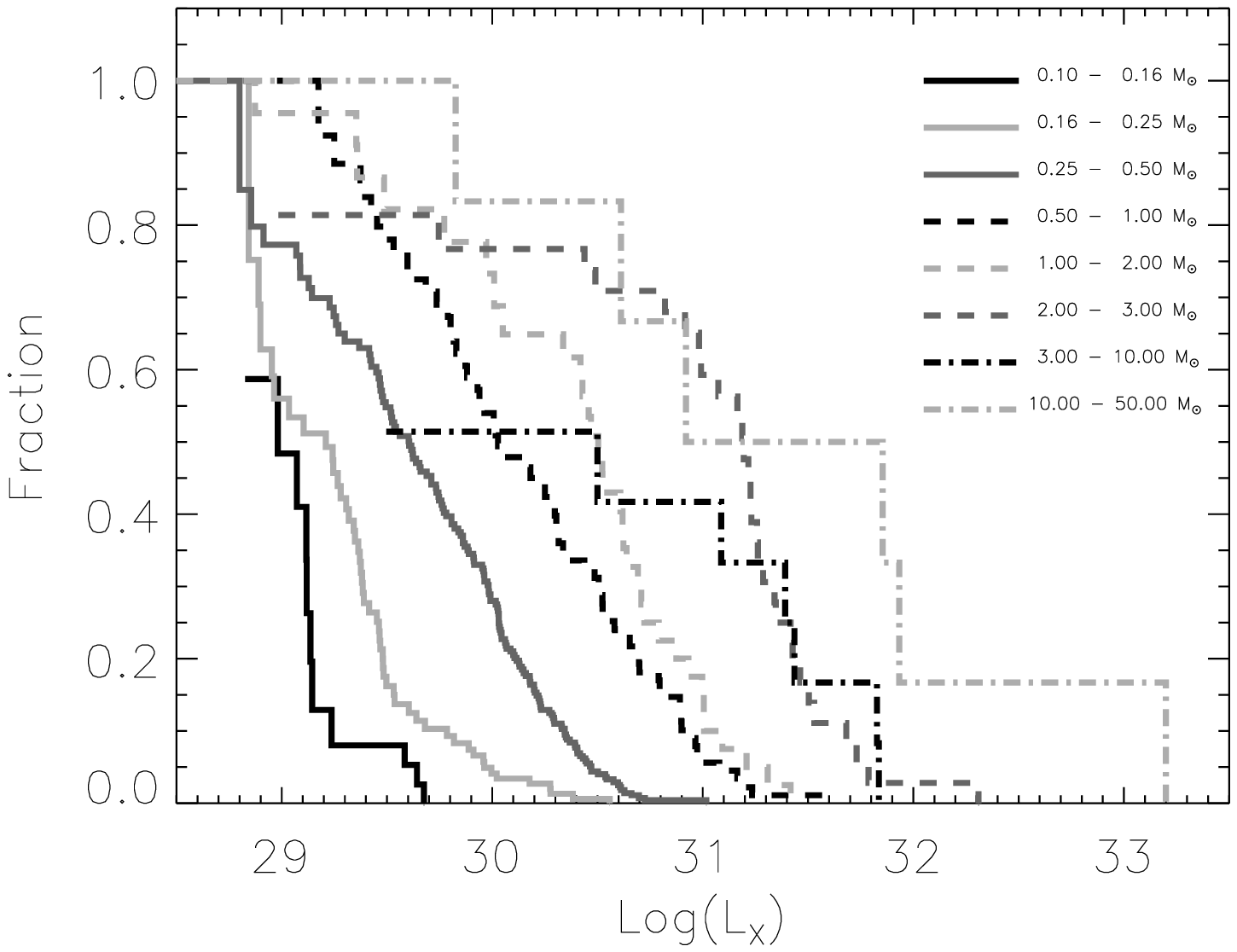,width=15cm}}
\caption{Maximum likelihood X-ray luminosity functions of the ONC {\em optical sample},
for eight ranges of mass from 0.1 to 50 $M_{\odot}$. Note that
for some distributions, low luminosity tails are not probed due to
insufficient sensitivity: the faintest stars in
such bins were not detected.\label{fig:LXvsMa}}
\end{figure}

\begin{figure}[!h!]
\centerline{\psfig{figure=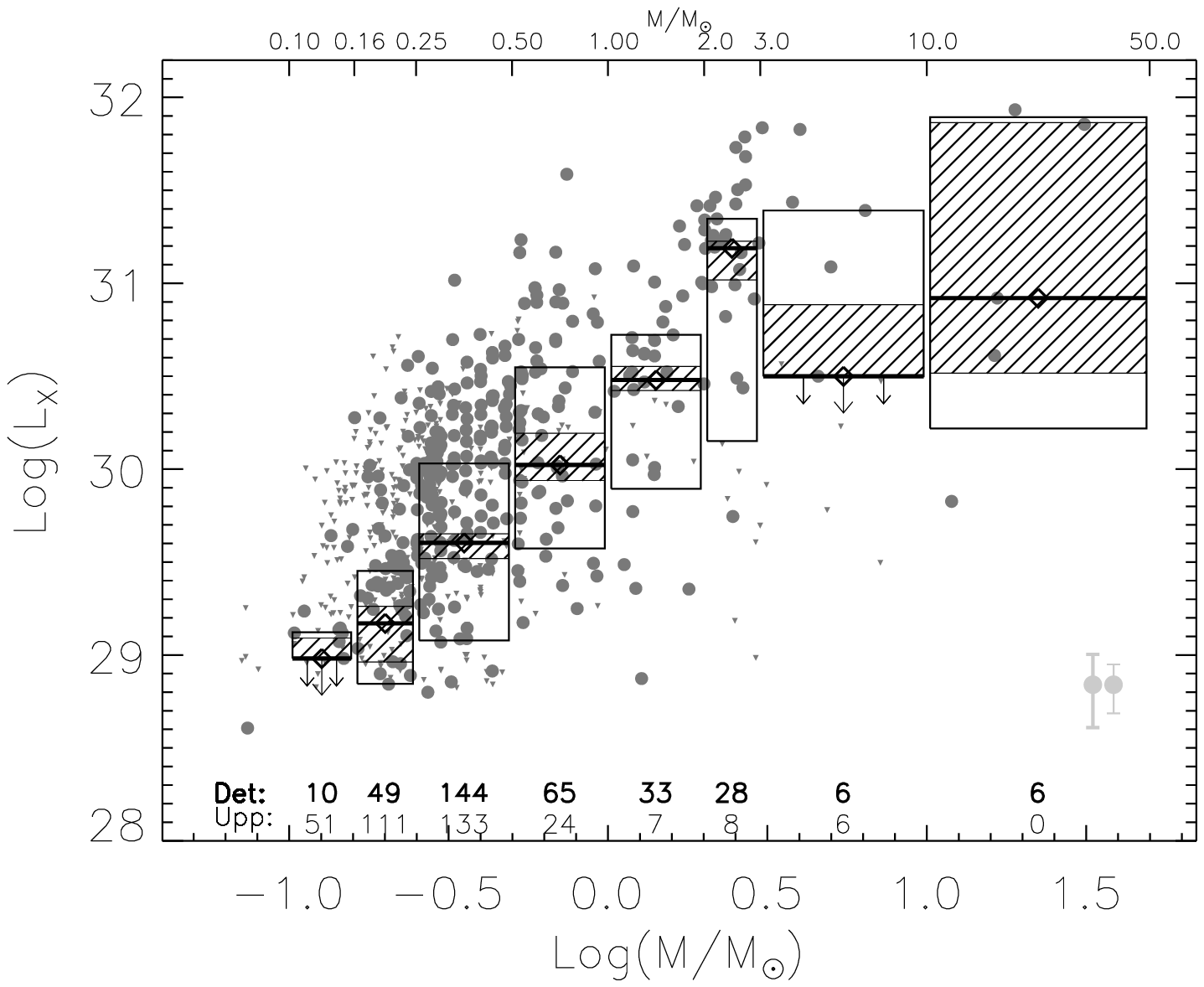,width=15cm}}
\caption{Scatter plot of $Log(L_X)$ vs. $Log(M/M_{\odot})$ for the {\em optical
sample}, with filled circles denoting detections and small downward-pointing
triangles, upper limits. Typical $1\sigma$ and 50\% uncertainties are given by
thick and thin error bars on two illustrative data points at lower right. Also
indicated are medians (diamonds centered on thick segments) with uncertainties
(shaded boxes), and the 25\% and 75\% quantiles (thin-lined boxes). Upper
limits corresponding to the lowest detection in a given mass range are
substituted for undefined quantities: a diamond with downward-pointing arrow
indicates an upper limit to the median; two small downward-pointing arrows at
box bottom indicate an upper limit on the lower error in median and/or the
lower quartile. Numeric values above abscissa for each box indicate numbers of
detections (``Det.'') and upper limits (``Upp.'')  contributing to the XLF
calculation. \label{fig:LXvsMb}} 
\end{figure}

Figure \ref{fig:LXvsMa} shows Maximum Likelihood X-ray luminosity functions
(XLFs) for ONC members in eight mass bins, while Figure \ref{fig:LXvsMb} shows
a scatter plot of $L_X$ vs. mass with characterizations of the XLFs superposed
for the same mass bins (see caption). In Figure \ref{fig:LXvsMb} a direct
correlation between median $L_X$ and mass can clearly be seen for $M \lesssim 3
M_{\odot}$, while in the mass range from 3 to 10 $M_{\odot}$ (where 6 of 12
stars have luminosities below detection threshold) the median $L_X$ drops
significantly and only an upper limit can be estimated.  For massive O-type
stars with $M \gtrsim 10~M_{\odot}$, $L_X$ rises again.

Scenarios in which a dynamo drives X-ray activity would interpret
this break at $\sim$3~$M_{\odot}$ as the point at which stars of the
ONC's age ($\sim 1 Myr$) become fully radiative and the dynamo stops
functioning. This interpretation is indeed supported by the stellar
structure evolutionary tracks of \citet[][ hereafter SDF]{sie00}.

The drop in $L_X$ at $\sim$3~$M_{\odot}$ is even more pronounced than shown in
Figures \ref{fig:LXvsMa} and \ref{fig:LXvsMb} if we take stellar multiplicity
into account.  The fact that most secondaries will have significantly lower
masses than their primaries, together with the mass~--~$L_X$ relation, means
that multiplicity effects on the XLFs can generally be expected to be rather
small.  But if primaries in the $3-10~M_{\odot}$ bin have very little intrinsic
X-ray emission, we will detect only secondary emission.  Indeed, we observe
that the XLF for $3-10~M_{\odot}$ mass stars crosses those for stars of lower
mass and is determined for about half the bin population. This lends indirect
support to the hypothesis that emission for $3-10~M_{\odot}$ mass stars is
mainly due to {\em unresolved} secondaries.

\begin{figure}[t]
\centerline{\psfig{figure=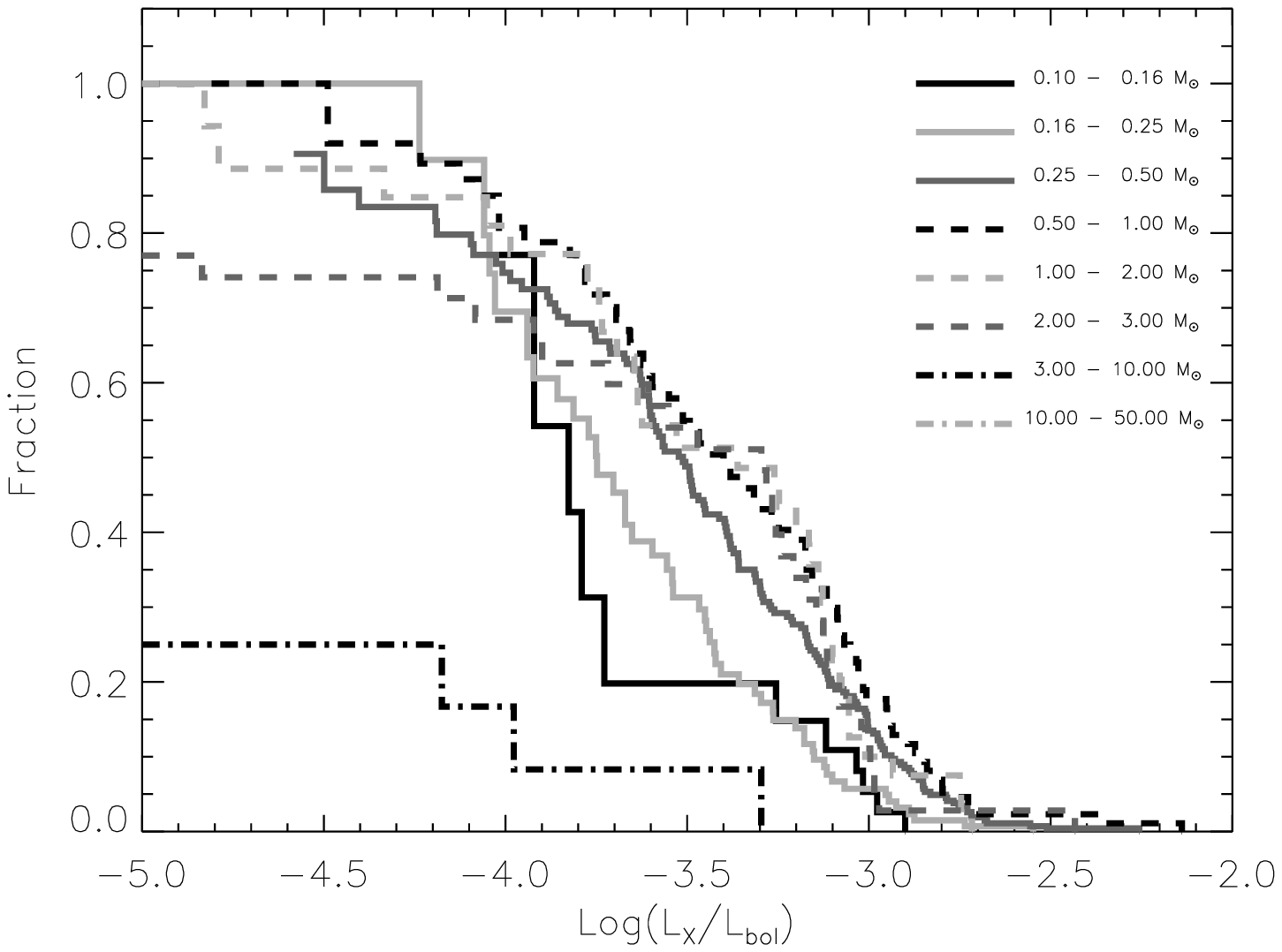,width=15cm}}
\caption{Maximum Likelihood distribution functions for $L_X/L_{bol}$, similar to Figure \ref{fig:LXvsMa}.
\label{fig:LXLBvsMa}}
\end{figure}

\begin{figure}[!h!]
\centerline{\psfig{figure=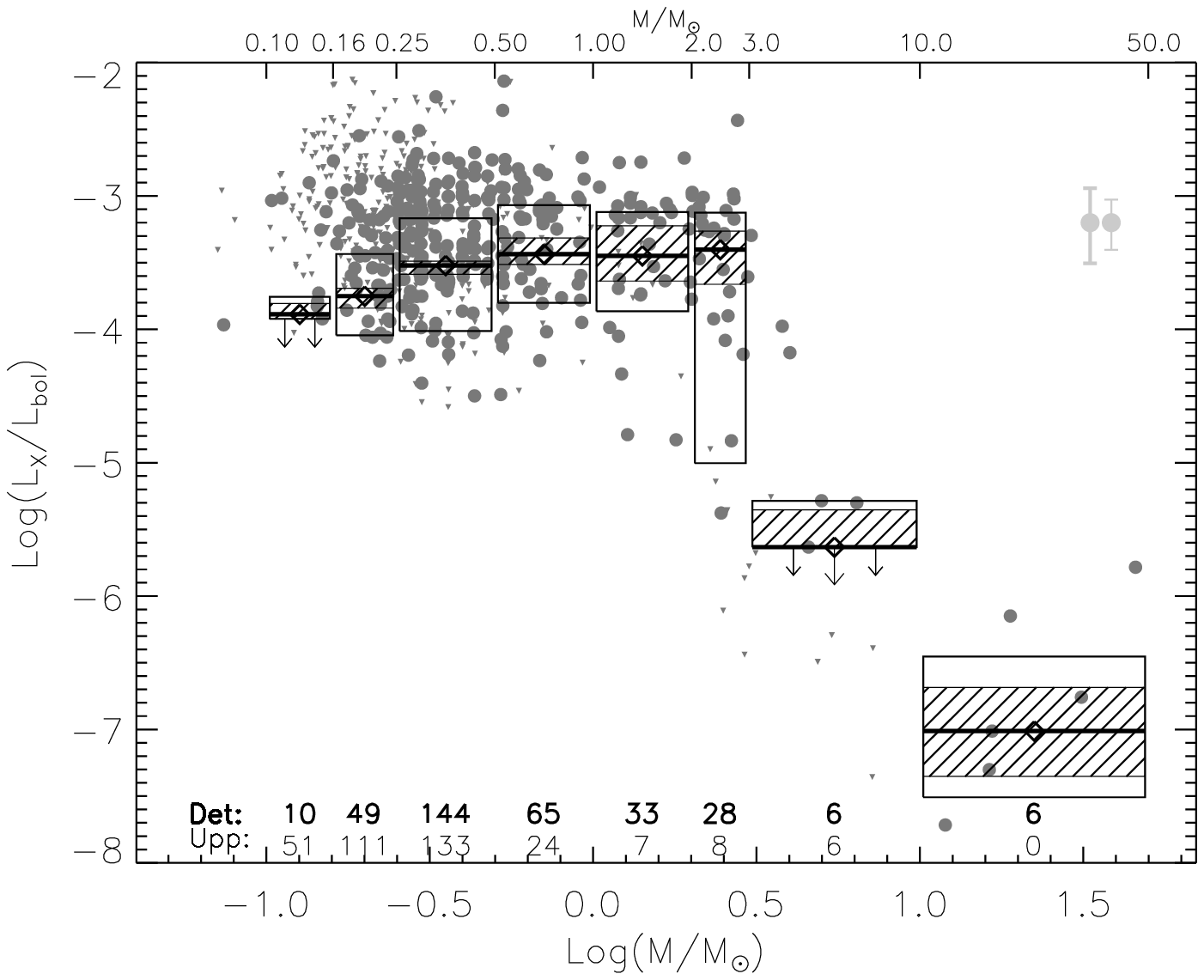,width=15cm}}
\caption{Log plot of $L_X/L_{bol}$ vs. mass, using conventions detailed in caption of Figure \ref{fig:LXvsMb}.
\label{fig:LXLBvsMb}}
\end{figure}

Figure \ref{fig:LXLBvsMa} shows distribution functions for $L_X/L_{bol}$, in
the same mass bins as defined for Figure \ref{fig:LXvsMa}, while Figure
\ref{fig:LXLBvsMb} follows the format of Figure \ref{fig:LXvsMb} to show
$L_X/L_{bol}$ vs. mass (logarithmically). The drop in activity at $3 M_{\odot}$
is even more striking with $L_X/L_{bol}$ as an indicator: more than 2 orders of
magnitude (note that the median value in the $3-10 M_{\odot}$ range is an upper
limit). It is also interesting to note that the upper envelope of 
$log(L_X/L_{bol})$ for stars with $M \lesssim 3 M_{\odot}$ is compatible
(within uncertainties) with the level of $\sim -3$, i.e. the {\em saturation
level} found for both rapidly-rotating main sequence stars \citep{vil87} and 
low mass stars in several other star formation regions
\citep[e.g.,][]{alc97,fla00a}. This is also apparent from Figure
\ref{fig:LXLBvsMa}. No satisfactory theoretical explanation for the physical
nature of this saturation has been found to date. Our analysis quantifies
typical $L_X/L_{bol}$ values in terms of the median and the scatter of points
and finds somewhat lower values than the canonical -3 value, indicating either
a different saturation level or perhaps the presence of a mix of saturated and
non-saturated stars. There also seems to be a relationship between median level
and stellar mass, with the lowest mass stars (e.g., those in the
0.10-0.16$M_{\odot}$ bin)  having a significantly lower median value of
$L_X/L_{bol}$ than those of higher mass stars. The range of the variation
between the 0.10-0.16$M_\odot$ and the 0.5-1.0$M_\odot$ bins is 0.45 dex,
larger than our estimated uncertainty in {\em individual} $L_X/L_{bol}$ values
(cf. Paper~I). A viable theoretical explanation of activity in the PMS phase
will have to account for this mass dependence of either the fraction of
saturated vs. unsaturated stars or of the saturation phenomenon itself.

To conclude our study of X-ray activity vs. stellar mass, we now wish to address
the X-ray properties of the significant fraction of undetected very low mass
stars. More specifically we wonder whether non detected stars are simply
located at the bottom of the $L_X$ distributions shown in Figure
\ref{fig:LXvsMa} or whether they constitute distinct stellar populations with
significantly lower X-ray luminosities.  Appendix \ref{app:nodet} describes how
we used our {\em Chandra} HRC data to derive (or constrain) mean X-ray
luminosities for mass segregated groups of objects, some or all of which are
individually undetected. The method measures the flux of a ``composite''
source by summing HRC photons detected at the optical positions of individually
undetected objects comprising the ``composite.''

Figure \ref{fig:BD} shows the logarithm of average $L_X$, both for the
whole reference sample\footnote{Note that this reference sample contains fewer
objects than the complete {\em optical sample}. This is due to a number of
constraints imposed when selecting objects to which our method was applied (cf. Appendix A).}
and for undetected objects in several mass bins. The height
of the boxes reflect the uncertainty in the count-rate to flux conversion
factors for the ``composite'' source (see Appendix \ref{app:nodet}). The number
of summed objects in each mass bin is reported in the upper part of the
figure, and for reference, the median $L_X$ from Figure \ref{fig:LXvsMb} is
also shown.  As a ``sanity check'' we compared the average $L_X$ computed with
the method described in Appendix \ref{app:nodet} for the whole reference sample
of detected and undetected objects with the average $L_X$ computed from the
maximum likelihood luminosity functions (not shown in the figure). We find the
two results consistent, lending support to this analysis. Note that the
discrepancy of mean and median $L_X$ in Figure \ref{fig:BD} is indeed due to a
real difference between the two indices and indicates an asymmetry in the $L_X$
distributions.

From Figure \ref{fig:BD} we therefore draw the conclusion that undetected stars
in the stellar samples ($0.1 < M/M_{\odot} < 0.5$) have an average $L_X$ well
below our sensitivity limit (note that the 0.1-0.16$M_{\odot}$ bin has {\em only} an
upper limit). This could either mean that the XLFs have very long
low-luminosity tails or indicate the existence of a stellar population with
quiescent/suppressed magnetic activity.

The same qualitative results found in this section are also obtained if the
evolutionary tracks of \citet{dan97} are substituted for those of SDF in
estimating masses. Similarly, if we exclude the few {\em optical sample} stars
with unknown astrometric membership (cf. Paper I) from our analysis, the
results also remain unchanged.

\subsection{Brown Dwarfs \label{sect:bd}}

Our {\em optical sample} contains 15 substellar mass objects ($0.008 < M <
0.08~M_{\odot}$) all of which were studied through IR spectroscopy by
\citet{luc01}. Only one of the brown dwarfs (BD), with an estimated mass of
$M=0.074~M_{\odot}$, is detected in our HRC data\footnote{The brown dwarf
status of this object is uncertain. The quoted mass is taken from
\citet{luc01}. However, an independent mass estimate  from the optical data of
\citet{hil97} and the SDF tracks (cf. Paper~I) gives a larger value of
$0.11~M_{\odot}$.}. Our XLF analysis therefore cannot be extended to such low
masses.  This poses the question of why so few brown dwarfs are detected: Are
BDs subluminous in X rays with respect to our mass~--~$L_X$ relation, or do
they fall on this relation but below our limiting sensitivity level, as a
linear extrapolation of the mass~--~$L_X$ relationship in Figure
\ref{fig:LXvsMb} indeed suggests.

\begin{figure}
\centerline{\psfig{figure=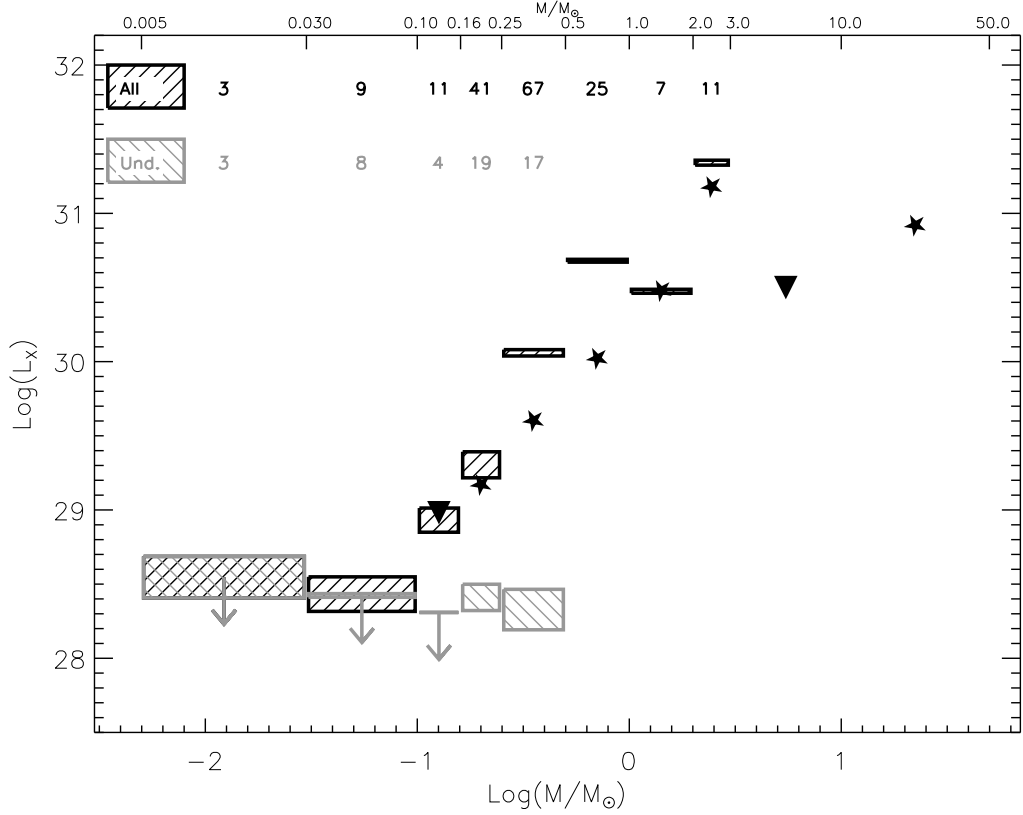,width=15cm}}
\caption{Mean and median $L_X$ vs. mass for a subsample of {\em optical sample}
stars and brown dwarfs in different mass bins (see text for details).  Hatched
boxes (small boxes appear as mere segments) represent mean $L_X$ derived for
the whole sample (black boxes) and for the subset of undetected objects (gray
boxes); see Appendix \ref{app:nodet}.  Box widths reflect mass bin sizes, while
heights reflect conversion factor (CF) uncertainties by indicating values
obtained with the mean and the median CFs of the individual objects within the bin. Stars
and upper-limit triangles represent $L_X$ medians from Figure
\ref{fig:LXLBvsMb}. Rows of numeric values at top give numbers of objects used
to compute averages for each mass bin.\label{fig:BD}}  
\end{figure}

In order to investigate this question, we once again applied the method of
Appendix \ref{app:nodet}: the two substellar mass bins in Figure \ref{fig:BD}
show our results. The lowest mass bin contains just 3 undetected objects and
yields only an upper limit.  Nine objects lie in the {\em brown dwarf} bin
($0.03 < M_{\odot} < 0.1$)\footnote{Although this mass range nominally extends
above the BD limit, the most massive object in this bin has a mass of only
$0.08~M_{\odot}$.} and our method yields a positive detection of the
``composite'' source with high confidence\footnote{An integrated flux of
$\approx$40.1 photons in a total effective exposure time of 553~ksec is
measured. As discussed above, one of these 9 objects is detected individually,
with 16.2 extracted counts. If we exclude this object of uncertain BD status,
the resulting ``composite'' source falls slightly below our detection threshold
($3\sigma$), with its probability of being due to background fluctuations at
$0.28\%$ ($2.8\sigma$). Figure \ref{fig:BD} indicates it as an upper limit
corresponding to less than 25.1 photons accumulated over an effective exposure
time of 491~ksec. Had we instead accepted it as a positive detection, an
accumulated photon count of 23.9 would imply an X-ray luminosity virtually
identical to the upper limit value.}.  Figure \ref{fig:BD} indicates that: 1)
brown dwarfs in the ONC do {\em not} depart from the $L_X$ - mass relationship
we observe for higher mass stars, and 2) the mean BD X-ray luminosity is of the
order of $10^{28.5}$ erg$\cdot$s$^{-1}$.  On average this is higher than the
recent results obtained for BDs in IC~348 by \citet{pre01}, but lower than the
$L_X$ reported by \citet{ima01} for one detected BD and one BD candidate in the
young $\rho$ Ophiuchi cloud.

\subsection{Stellar age \label{sect:LxvsAge}}

\begin{figure} 
\centerline{\psfig{figure=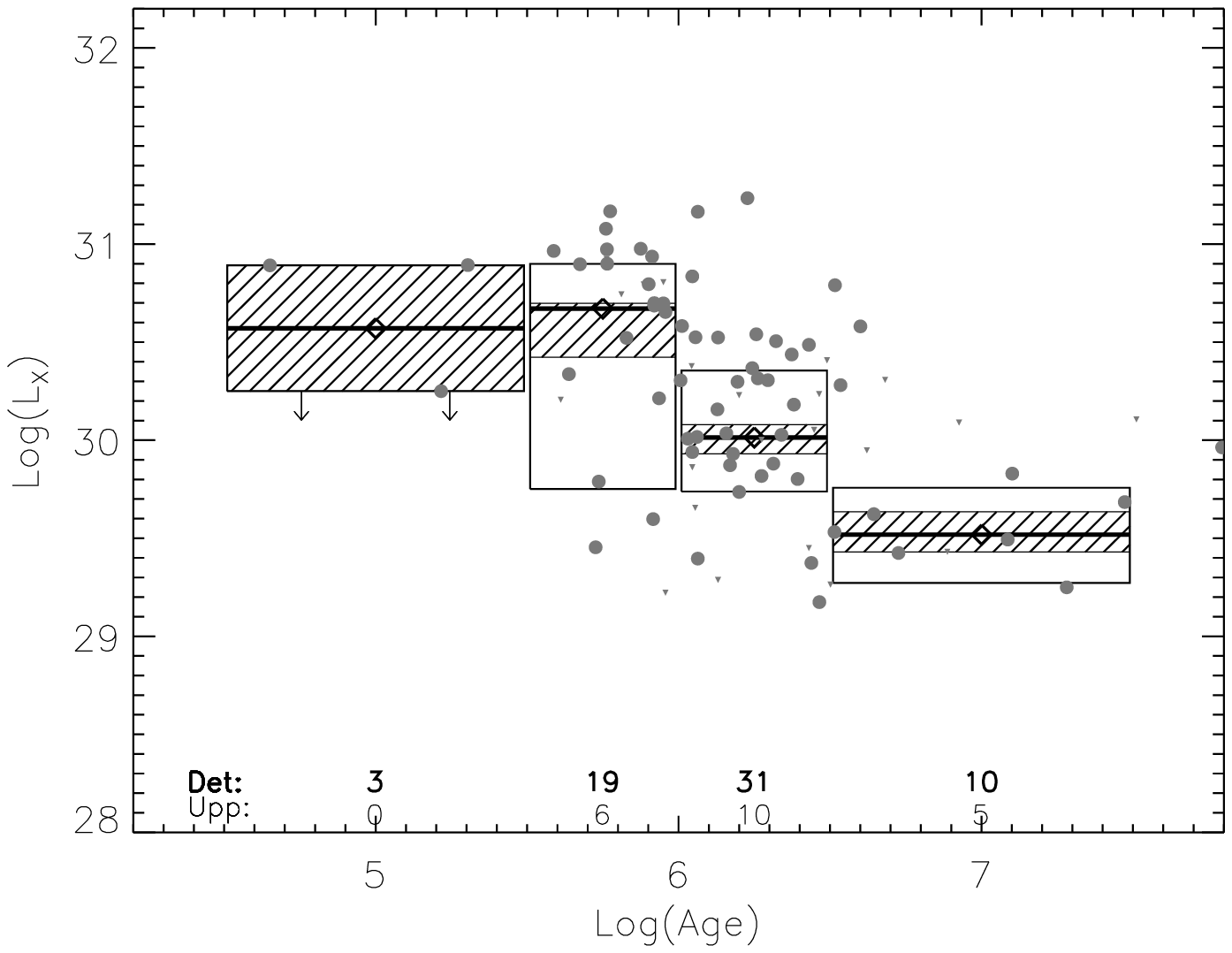,width=14cm}}
\caption{$Log(L_X)$ vs. Log(Age) for stars of mass 0.5-1.0~$M_{\odot}$; symbols same as Figure \ref{fig:LXvsMb}.
\label{fig:LXvsA0a}} 
\centerline{\psfig{figure=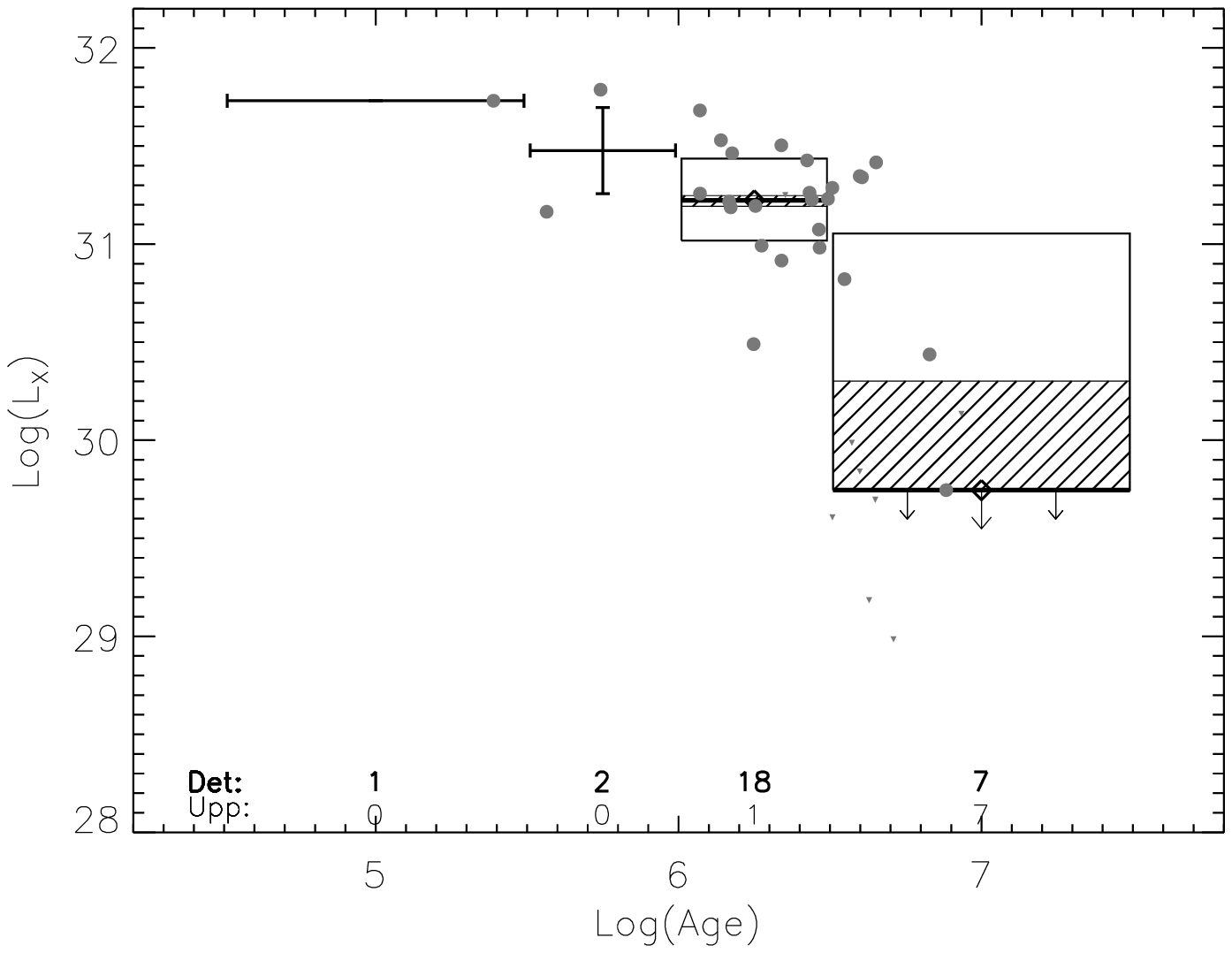,width=14cm}}
\caption{$Log(L_X)$ vs. Log(Age) for stars of mass 2.0-3.0~$M_{\odot}$; symbols same as Figure \ref{fig:LXvsMb}. Medians for the two left-most bins have been replaced by averages.
\label{fig:LXvsA0b}}
\end{figure}

Although the relationship with mass explains a good deal of the scatter in
activity levels of our sample, the deviations from this trend are still
significant. The next stellar parameter we investigate in order to understand
this residual scatter is the age, as inferred from the SDF evolutionary tracks.
Figure \ref{fig:LXvsA0a} shows the $Log(L_X)$ vs. $Log(Age)$ plot for stars in
our sample with mass between 0.5 and 1.0 $M_{\odot}$. The median $L_X$ appears
to decrease with increasing age, but we must exercise special care in the
interpretation of this trend: there is a widespread concern that the age spread
indicated by the position of low mass stars still on the Hayashi tracks may be
largely, or even entirely, due to an artificial spread in the bolometric
luminosities \citep[cf.][]{har01}. The sources of error that may contribute to
uncertainties in $L_{bol}$ and consequent age are indeed numerous and include
uncertainties in: adopted distance, spectral type and extinction, photometric
variability, unresolved companions, and accretion luminosity. On the other
hand, if one were to concede a real age spread in the ONC (plausible as star
formation is still taking place at the present day), stellar ages might carry a
statistical and relative significance, though still uncertain on an individual
basis: stars that we place lower on the Hayashi tracks may be older, on
average, than those that we place higher. This latter hypothesis is supported by the time evolution of ONC stellar radii recently observed by \citet{rho01}. Although our correlation of $L_X$
with age also seems to argue in favor of a real age spread, it could also be an
artifact of an interrelation between inferred $L_X$ and $L_{bol}$, and the
matter remains open. Indeed the trend in Figure \ref{fig:LXvsA0a}
corresponds to a constant ratio of X-ray to bolometric luminosity at different
ages and can be interpreted equally well in two different ways: 1) Stars of
equal mass have equal bolometric and X-ray luminosities, but there are effects
that act in exactly the same way on both $L_X$ and $L_{bol}$; 2) These stars
are saturated and stay saturated through their contraction on the Hayashi
tracks, thus keeping $L_X/L_{bol}$ constant.

Figure \ref{fig:LXvsA0b} presents a plot of $Log(L_X)$ vs. $Log(Age)$ for
stars in the $2-3~M_{\odot}$ mass bin and reveals an age dependence and a sudden
drop of $L_X$ at $Log(Age) \sim 6.5$ (a similar drop is also seen in
$L_X/L_{bol}$).  This is just the age at which a $2.5~M_{\odot}$ star
dissipates its convective envelope according to the SDF models, lending further
evidence for the convection~--~activity connection found in \S
\ref{sect:LxvsMass}.  Figure \ref{fig:LXvsA0b} also supports the existence of a
real age spread in the ONC. Contrary to the situation in the
$0.5-1.0~M_{\odot}$ mass range, evolution of these higher mass stars occurs in
temperature as well as in luminosity for $Log(Age)~\gtrsim~6.0$.  The spread in
age therefore cannot be due to a spurious spread in $L_{bol}$, as may be the case
in the lower mass bin. 

\begin{deluxetable}{cl|rrrrrrr}
\tabletypesize{\small}
\tablewidth{0pt}
\tablecaption{Detections and Upper Limits for mass/age ranges }

\startdata
\hline
& \multicolumn{7}{c}{Mass [$M_\odot$]}  \\ 
\multicolumn{2}{l}{Log(Age) [yr.]}   
&0.1-0.16&0.16-0.25&0.25-0.50&0.5-1.0&1.0-2.0&2.0-3.0&3.0-10.0   \\
\hline

4.5-5.5   &Det.  &  1&	14&	 13&	  3&	  0&	  1&	 0\\
    "     &Upp.  &  6&	12&	  6&	  0&	  0&	  0&	 0\\
 
5.5-6.0   &Det.  &  3&	 7&	 27&	 19&	  1&	  2&	 3\\
    "     &Upp.  & 13&	16&	 18&	  6&	  0&	  0&	 1\\
 
6.0-6.5   &Det.  &  3&	21&	 95&	 31&	 12&	 18&	 2\\
    "     &Upp.  &  8&	43&	 80&	 10&	  3&	  1&	 3\\
 
6.5-7.5   &Det.  &  3&	 7&	  7&	 10&	 20&	  7&	 0\\
    "     &Upp.  & 24&	39&	 27&	  5&	  4&	  7&	 0\\
\enddata
\label{tab:n_star_ma}
\end{deluxetable}
   
\begin{figure}[!t!]
\centerline{\psfig{figure=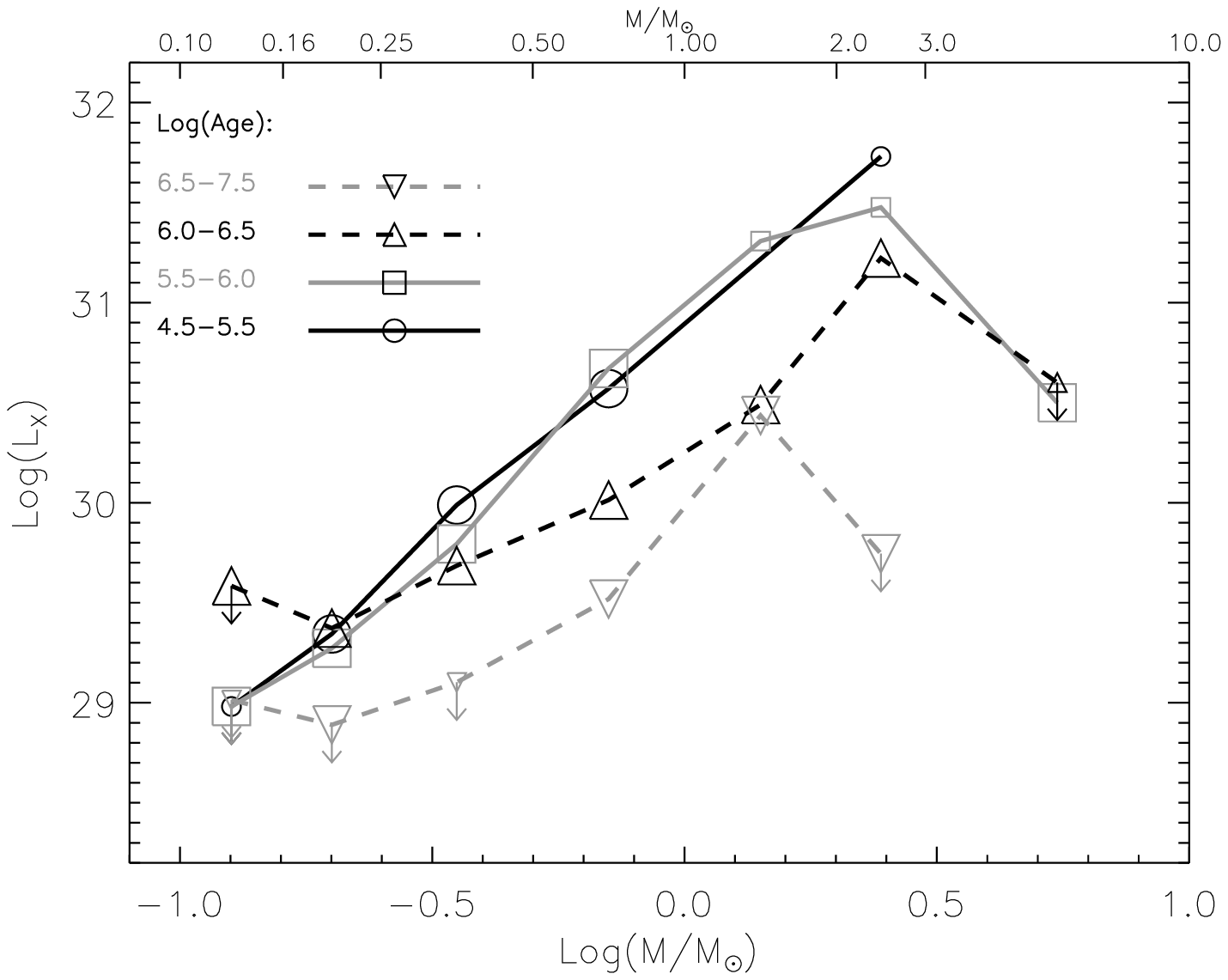,width=14cm}}
\caption{$Log(L_X)$ vs. $Log(M/M_{\odot})$ in four age ranges. Black- and
gray-colored symbols, connected with different-styled lines, give results for
age ranges designated by the legend at upper left. Large symbols indicate median
$Log(L_X)$ values from Maximum Likelihood distributions in each mass range;
small symbols indicate mean $Log(L_X)$ for those mass/age bins with $< 4$
detections. Downward-pointing arrows indicate upper limits on medians or means.
\label{fig:LXvsM4}}
\end{figure}

\begin{figure}[!h!]
\centerline{\psfig{figure=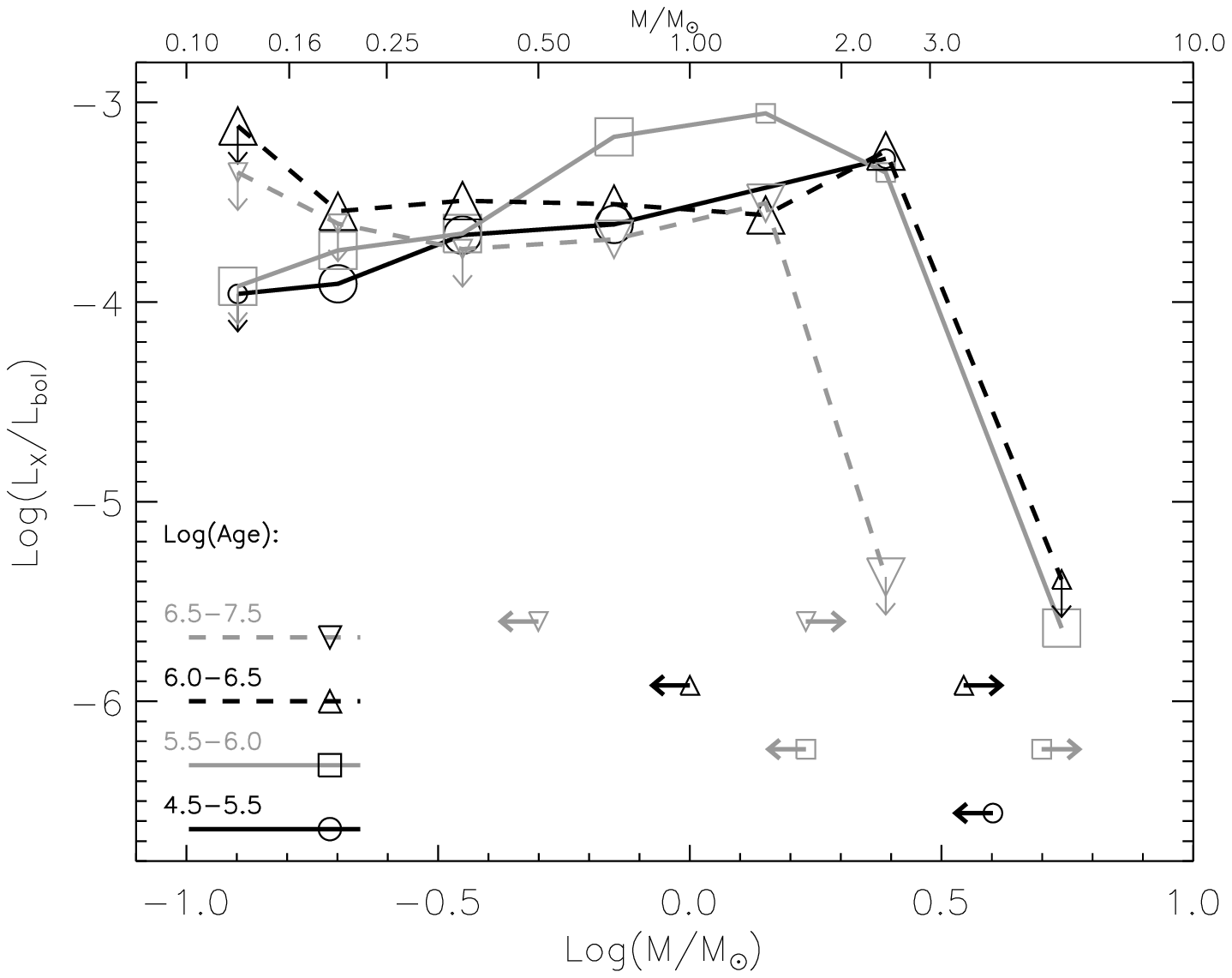,width=14cm}}
\caption{ $Log(L_X/L_{bol})$ vs. $Log(M/M_{\odot})$ in four age ranges, with
symbols and lines same as Figure \ref{fig:LXvsM4}. Lower portion of plot (right
of legend) indicates masses at which the structure of stars of $Log(Age)$
are expected to be fully convective (leftward arrows) and fully
radiative (rightward arrows), according to SDF evolution.\label{fig:LXLbvsM4}}
\end{figure}

Figures \ref{fig:LXvsM4} and \ref{fig:LXLbvsM4} show trends of median $L_X$
and $L_X/L_{bol}$ as a function of mass in each of four age bins. Care must be
taken in interpreting these plots because many of the mass/age bins for which
medians are computed are sparsely populated (numbers of detections and upper
limits are given in Table \ref{tab:n_star_ma}) and some points are subject to
rather large uncertainties. In those bins with less than four detections, we
show means computed from the XLFs (instead of medians), and both medians and
means are replaced by upper limits in some cases. From left to right in the
lower portion of Figure \ref{fig:LXLbvsM4} we show mass ranges over which stars
of each age bin are fully convective, partially convective, and fully radiative
according to SDF tracks.  We can draw several conclusions from this presentation:
1) For all ages, $L_X$ increases with mass for low masses and then
suffers a break at higher mass; 2) The mass at which this break occurs appears to
depend on the age, being lower for the oldest stars in our sample; 3) There is
good agreement between the mass at which the break occurs and the mass at which
stars in that age bin become fully radiative; 4) The increase of $L_X/L_{bol}$
with mass appears again at the low mass end, at least for $Log(Age) \lesssim
6.0$.


\subsection{Accretion/disk indicators: Ca~II \label{sect:CaII}}

\begin{figure}[!t!]
\centerline{\psfig{figure=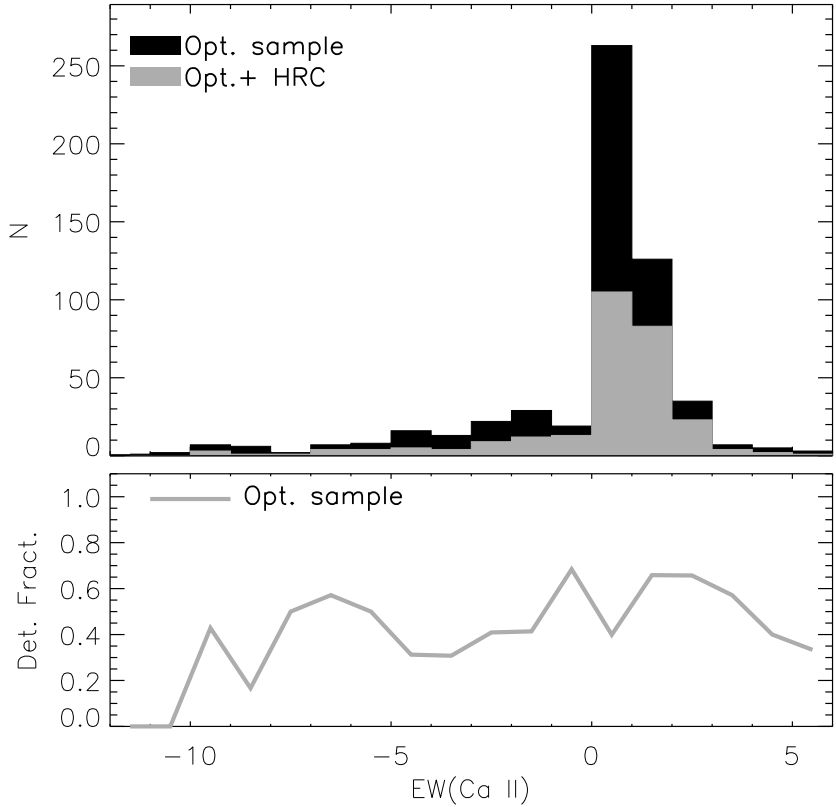,width=12cm}}
\caption{Upper panel: Distribution of Ca II ($\lambda = 8542\AA$) line equivalent width for  the {\em optical sample} and the subset of stars detected in our HRC data. Lower panel: detection fraction. \label{fig:caii_dist}}
\end{figure}

It has long been disputed whether the presence of circumstellar disks and/or
circumstellar accretion plays a role in determining the activity levels of PMS
stars. We are now in the position to address this problem in the ONC. We will
use the $\lambda = 8542\AA$ equivalent width ($EW$) of the Ca~II triplet as an
indicator of disk accretion. In main sequence and weak-line T Tauri stars,
where accretion is absent, this line is seen in absorption with $EW \sim 3$, a
value that depends only weakly on spectral type  \citep[cf. discussion
in][]{hil98a}. Accretion has the effect of filling in the line and the $EW$ is
then related to the mass accretion rate \citep{muz98}. The observational values
we will use are those obtained through $5-8\AA$ resolution spectroscopy by
\citet{hil97} for a large fraction of our {\em optical sample} (see Table 5 of
Paper~I). Figure \ref{fig:caii_dist} (upper panel) shows the distribution of
the Ca II $EW$ both for our {\em optical sample} and for the subset of HRC
detected stars. The detection fraction as a function of $EW$ is shown in the
lower panel. 

In order to make the distinction between accreting and non accreting stars more
evident and to minimize the effects of observational uncertainties\footnote
{\citet{hil98a} quote $EW$ errors of $\pm 0.5$.}, we divided our stars in two
groups: stars whose Ca~II line is seen in absorption, with $EW($Ca~II$) > 1$,
and which therefore are inferred to have low or no accretion, and stars that
show the line in clear emission, with $EW($Ca~II$) < -1$, and which are
undergoing significant accretion.  We exclude from our analysis stars with $-1
< EW($Ca~II$) < 1$, about half the sample for which we have Ca~II $EW$
information (see Figure \ref{fig:caii_dist}).  As the majority of stars in the
ONC appear to possess a circumstellar accretion disk, judging from both Ca~II
line and IR photometric indicators \citep{hil98a,lad00,mue01}, we refer to
stars in the two selected groups as having {\em high} and {\em low}
circumstellar accretion and avoid using the terms ``classical'' or
``weak-line'' T Tauri stars.

\begin{figure}[t]
\centerline{\psfig{figure=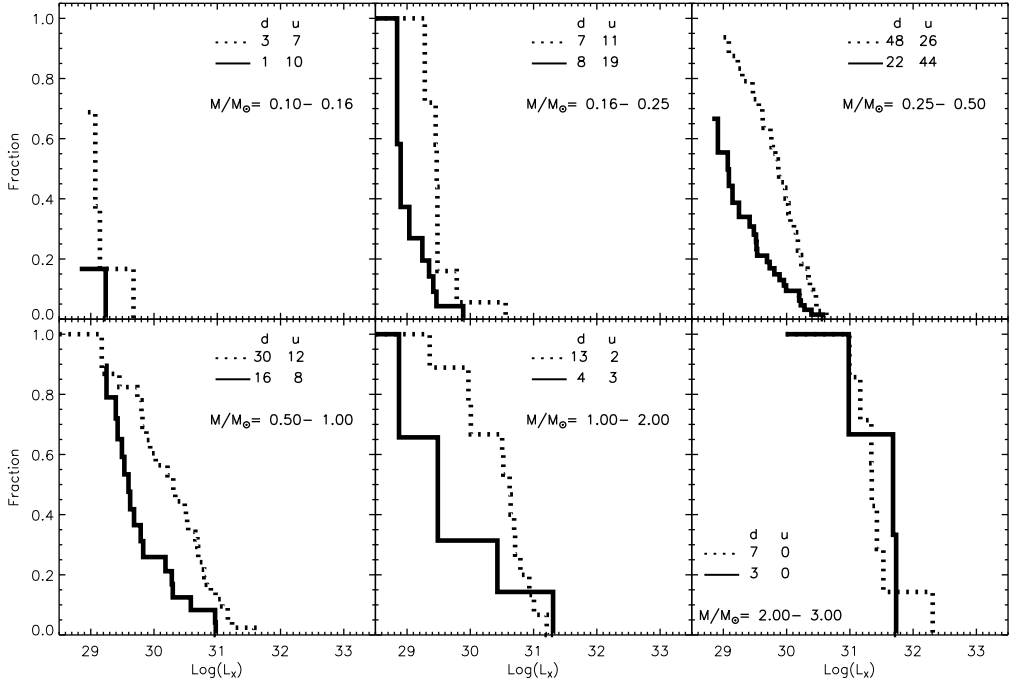,width=14cm}}
\caption{{\em Optical sample} XLFs for stars with high- and low-accretion (solid
and dashed lines, respectively). Panels give different mass ranges, as indicated
by legends. Legends also give numbers of detected (d) and undetected stars (u)
used for XLFs of high- and low-accretion subsamples. \label{fig:XLF_CW}}
\end{figure}
\begin{figure}[!t!]
\centerline{\psfig{figure=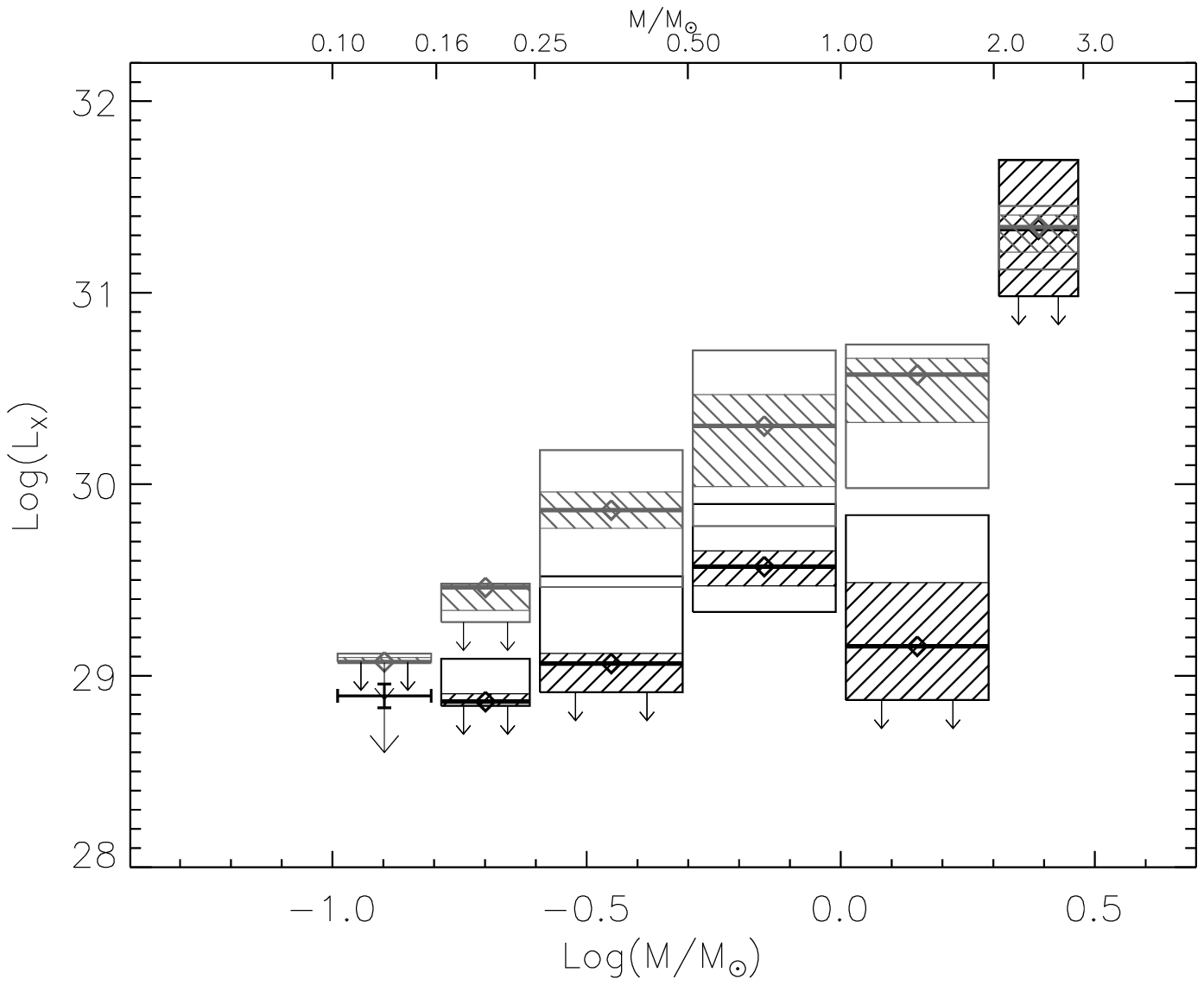,width=14cm}}
\caption{$Log(L_X)$ vs. $Log(M/M_{\odot})$ for high-accretion (black) and
low-accretion (gray) stars.  Symbols have same meaning as in Figure
\ref{fig:LXvsMb}, but with individual data points omitted for clarity. Because only three detections are found in the lowest mass bin of the high-accretion sample, we show an upper limit to the mean $L_X$ in place of the median. \label{fig:LXvsM_CWa}}
\end{figure}

In Figure \ref{fig:XLF_CW} we present X-ray luminosity functions for stars in
the six lowest mass ranges defined in \S \ref{sect:LxvsMass}, computed
separately for the two accretion-differentiated samples.  Figure
\ref{fig:LXvsM_CWa} shows median $Log(L_X)$ vs. $Log(M/M_{\odot})$ depicted in
the same format as Figure \ref{fig:LXvsMb} and separated into low- and
high-accretion subsamples.  For any mass bin with enough points, it is clear in
either representation that stars with high accretion have lower median $L_X$
than low-accretion stars.  The difference of the medians is about an order of
magnitude in the $0.25-2.0~M/M_{\odot}$ mass range. Two-population statistical
tests confirm the difference with high confidence (null-hypothesis
probabilities less than $10^{-4}$, for $M=0.25-0.50~M/M_{\odot}$). Similar
plots for $L_X/L_{bol}$ confirm the same result, also with high confidence:
low-accretion objects are more X-ray active that similar high-accretion
objects. In order to check that the dependence of activity on accretion is not
due to different mean ages/bolometric luminosities in the two subsets, we
repeated this analysis in several mass {\em and} age slices, similar to those
of Figure \ref{fig:LXvsM4}.  When we did this, we invariably confirmed that the
high-accretion stars are less X-ray active than the low-accretion ones. For any
mass and age bin where the two subsamples are well represented, we obtain
highly significant results (e.g., in the bin with $M/M_{\odot}=0.25-0.50$ and
$Log(Age)=6.0-6.5$, null-hypothesis tests give better than 99.96\% confidence
that the two accretion categories differ).

Differences in X-ray activity level aside, trends of mean activity with mass
seem to hold preferentially for low accretion stars.  The scatter in X-ray
activity levels in any given mass range also appears lower for low-accretion
stars than for the whole sample (cf. Figures \ref{fig:LXvsMb} and
\ref{fig:LXvsM_CWa}). A relationship of activity with mass is {\em not} apparent for
high-accretion stars, which also evidence larger spreads in their activity levels.

\section{The nature of unidentified sources  \label{sect:unid}}

Forty of our HRC sources remain unidentified with any optical/IR object. In
this section we will explore the nature of these sources.

\begin{figure}[!t!]
\centerline{\psfig{figure=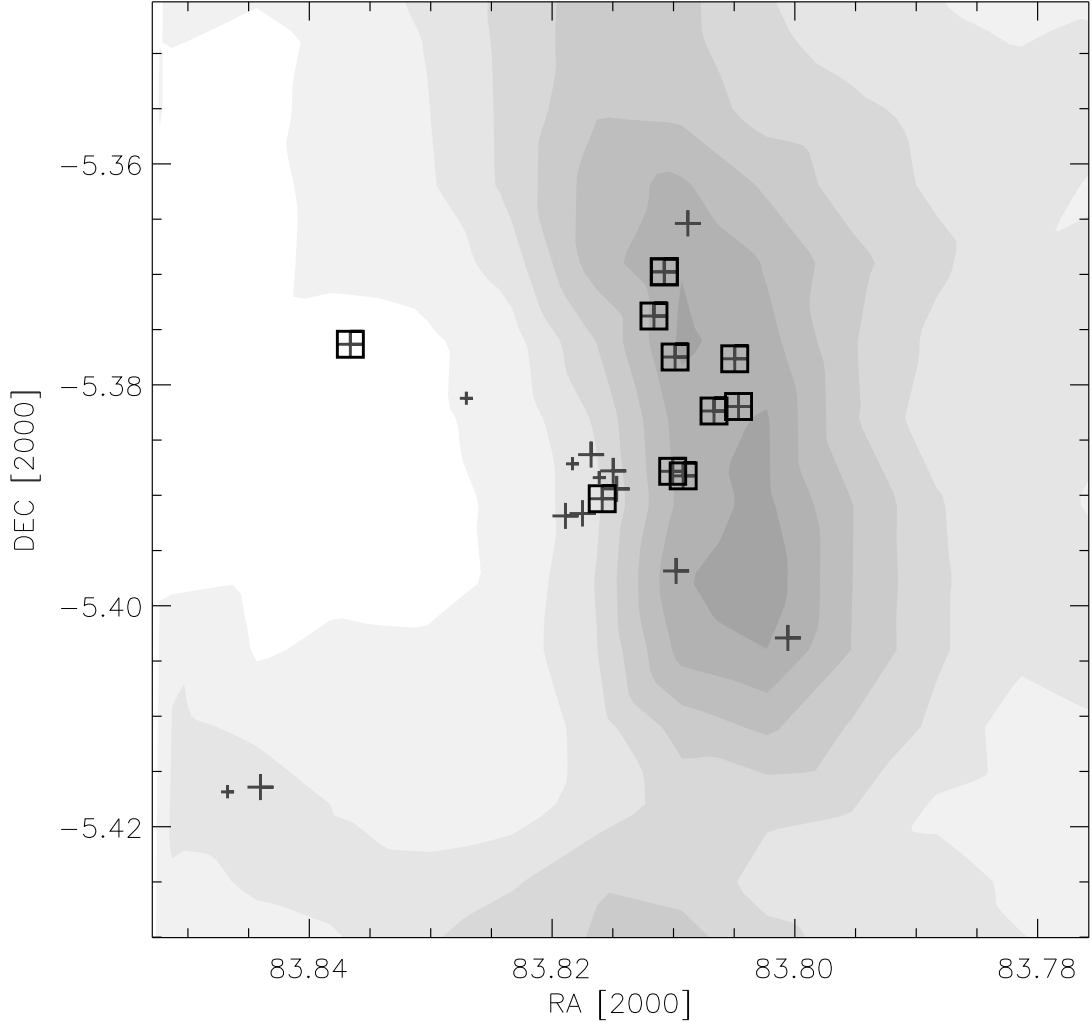,height=14cm}}
\vspace*{0.8cm}
\caption{Spatial distribution of ``non-IDs'' (pluses) in the inner $\sim
5'\times5'$ of the HRC field; squares indicate those previously detected by {\em
Chandra} \citep{gar00,sch01}. Shaded contours trace molecular cloud total
optical extinction (from \citealt{gol97}), with gray scale running from $A_V < 10$
(white) to $A_V > 70$ (dark grey) in steps of 10 magnitudes. \label{fig:noid_dist}}
\end{figure}

As noted in Paper~I (cf. \S 2.2), our X-ray detection method will yield about
10 spurious detections, and virtually all of these are expected to be
unidentified (Paper~I, \S 3.2) and have near-threshold SNRs.  We therefore have
$\sim 30~(=40-10)$ sources with no optical/IR identification ("non-IDs") that
are expected to be real and require some explanation\footnote{Of the 40
``non-IDs'' 14 were previously detected with {\em Chandra} by \citet{gar00}
and/or \citet{sch01}; we are therefore confident that these 14 X-ray detections
are {\em not} spurious. \label{ind_conf}}. Particularly puzzling is the fact
that most of them appear in the central $5'\times5'$ of the FOV (see Figure
\ref{fig:noid_dist}) where the $K$-band coverage is quite deep (limiting $K
\sim 17.5$). A first clue to their nature comes from their spatial distribution
and from the correlation with the \citet{gol97} $^{13}$CO data on the molecular
cloud total extinction, as shown in Figure \ref{fig:noid_dist}. If we exclude
four, likely-spurious\footnote{We note that the 26 (=40 - 14, see footnote
\ref{ind_conf}) unidentified detections that have not been confirmed have
extremely low SNRs (median SNR = 4.94, vs. 12.45 for those previously confirmed
by {\em Chandra}).} sources with $SNR < 4.8$, the median $V$-band total
extinction at the positions of the remaining sources is $\sim$63 mag, much
larger than that for all detected sources ($\sim$24 mag) in this central
region. This suggests that the "non-IDs" as a group are high-intrinsic X-ray
luminosity objects, which are either highly embedded in the cloud or which lie
in the background.

This conclusion is supported by our analysis of the hydrogen column density,
$N_H$, measured directly from X-ray data (cf. Appendix A of Paper~I).  Since
the ``non-IDs'' are largely low-SNR detections$^9$, however, the hardness
ratios derived from Advanced CCD Imaging Spectrometer (ACIS, \citealt{tow00})
spectra are subject to large uncertainties.\footnote{The hardness-ratio
approach suffers not only from low source statistics but also from the fact
that the low-energy hardness ratio saturates at high values of extinction (see
Paper~I, Appendix A).} Nevertheless, the fact that the low-energy portion of
their spectra are extremely underpopulated with respect to higher energies is
consistent with the hypothesis of high extinction: on average, ``non-ID''
hardness ratios suggest optical extinction $\gtrsim 30$~magnitudes.

Having established that a majority of our unidentified sources are associated
with highly absorbed objects, we next consider whether or not the hypothesis
that they are high-intrinsic luminosity, embedded cluster members is compatible
with the observational constraint that they are not detected in the deep
$K$-band survey. We can derive an empirical relationship between stellar mass
and intrinsic, non absorbed, $K$-band magnitude\footnote{There is significant
scatter in $K$ for this relation, probably due both to variable IR excesses
induced by the presence of circumstellar disks and to a spread in stellar
ages.} by considering the two quantities for ONC members with low measured
optical extinction ($A_V < 1.0$).  We then derive, as a function of mass, the
minimum $K$-band extinction needed to make these sources fall below the
$K$-band sensitivity threshold of $K= 17.5$. Converting this minimum $A_K$ to a
minimum $N_H = 9.3 A_K \cdot 2\cdot~10^{21} \ cm^{-2}$ and assuming a detection
threshold of 10 photons, we then find the minimum $L_X$ that stars emitting at
a given temperature and similar to those in our low-extinction sample should
have in order to be both detected in our HRC data and too absorbed to appear in
the deep $K$-band survey.

\begin{figure}[!t!]
\centerline{\psfig{figure=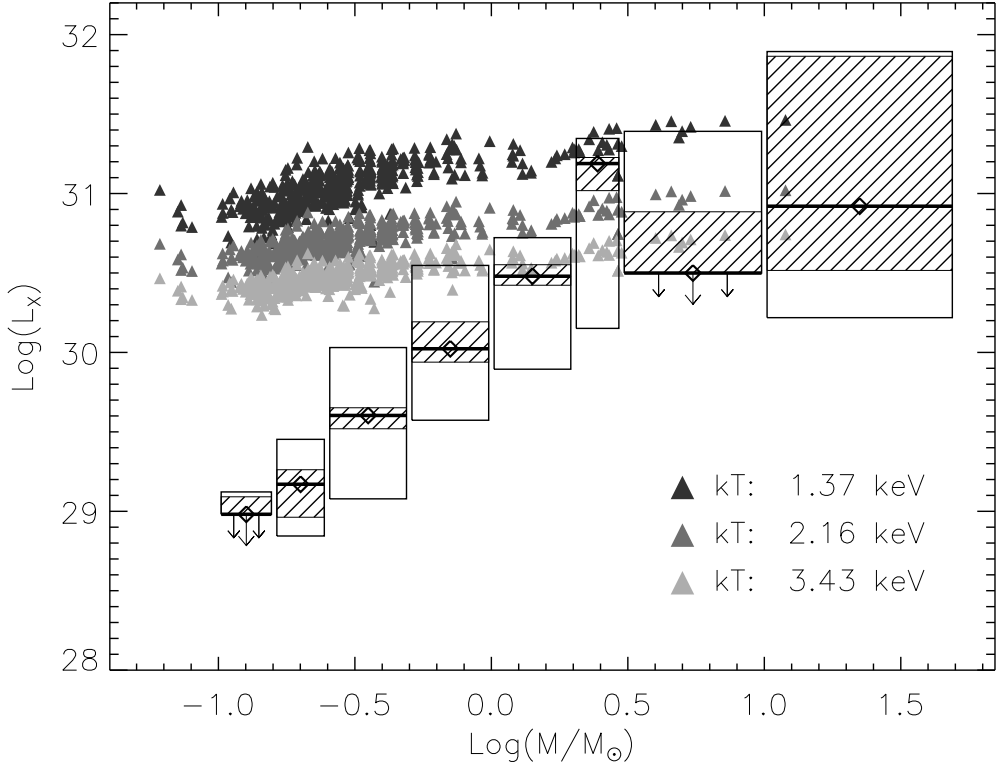,width=15cm}}
\caption{Log-log plot vs. mass of inferred minimum X-ray luminosity for objects
too faint to appear in the K-band survey of \citet{hil00}, yet X-ray luminous
enough for {\em Chandra} detection. Triangles give hypothetical lower limits for
stars in our {\em optical sample} with $A_V < 1$ (see text), computed for
different values of X-ray temperature as indicated by legend. For reference,
we also show the characterization of the XLF as presented in Figure~\ref{fig:LXvsMb}.
\label{fig:noid}} \end{figure}

Figure \ref{fig:noid} shows the result of this exercise for three different
assumed source temperatures within the range determined in Paper~I. Any source
that we detected without a $K$-band counterpart requires an X-ray luminosity in
excess of 10$^{30.3}$ erg$\cdot$s$^{-1}$. The large scatter in inferred $L_X$
values (for a given temperature) is due to the spread in the mass~--~$K$-band
magnitude relation.  We also show on the same plot the relation obtained in \S
\ref{sect:LxvsMass} between stellar mass and $L_X$ for the {\em optical
sample}.  We conclude that ``non-IDs'' may well be embedded ONC members but
must have masses larger than $\sim 1~M_{\odot}$, i.e. lower mass ONC members of
the same activity level as our {\em optical sample}, if detected in X rays,
{\em must} appear in the K-band survey and hence cannot be ``non-IDs.''

Thus we have established that the ``non-IDs'' are indeed affected by high
extinction and that their HRC count rates and $K$-band magnitude lower limits
are compatible with their being highly absorbed "high" mass PMS stars similar
to those of our {\em optical sample}. In order to confirm their ONC membership,
we must also exclude the possibility of their being background extragalactic
objects.

We can estimate the number of expected extragalactic X-ray sources by combining
information on the expected sky density of such objects, as a function of
unabsorbed flux \citep{gia01,toz01}, with our knowledge of the total extinction
due to the Orion Molecular Cloud \citep{gol97}. For this exercise we make the
following assumptions: 1) Extragalactic sources have power law spectra of
photon index $\Gamma$ between 1.3 and 1.5, compatible with values derived by
\citet{gia01}; 2) The sky density $\rho[F]$ of extragalactic objects as a
function of unabsorbed flux follows the descriptions of \citet{gia01} and
\citet{toz01}\footnote{They give two relations, $\rho_s(F_s)$ and
$\rho_h(F_h)$, for the sky density of objects with unabsorbed flux larger then
$F$, measured in {\em soft} (0.5-2.0~keV) and {\em hard} (2.0-10.0~keV)
spectral bands, respectively.}; 3) Observed photon flux is determined by
intrinsic flux and an absorption proportional to the molecular cloud total
optical absorption ($N_H=2\cdot 10^{21}A_V^{tot}$), as estimated from radio
data by \citet{gol97}; we use $CF(N_H)$ to denote the unabsorbed flux to HRC
count rate conversion factor computed with PIMMS, which depends on $N_H$, the
assumed source spectrum and the spectral band under consideration; and
4) We can detect sources above count rate thresholds, $CR_{th}$, computed as
the photon threshold (viz. 6, 8 or 10) divided by exposure time. With these
assumptions, the surface density of extragalactic sources seen through a given
hydrogen column  density becomes $\rho[CR_{th} \cdot CF(N_H)]$ and we can
compute the expected number of extragalactic sources detected in an area $S$
as:

\begin{equation}
\int_S \rho[CR_{th} \cdot CF(N_H)] \cdot d a
\label{eq:agn}
\end{equation}

\noindent
where $da$ is the infinitesimal area element and $N_H$ is a function of position
on the sky. Computing the integral (\ref{eq:agn}) for the given values of $\Gamma$, $CR_{th}$ and energy band (and therefore
$\rho[F]$), we find expected numbers of detectable extragalactic objects in the inner
$5'\times5'$ of our FOV ranging from 0.6 to 1.2. An
extragalactic nature for most of our unidentified sources is therefore
excluded, and most ``non-IDs'' can indeed be highly embedded ONC
members never observed with optical or near-IR instruments.

We can get a rough estimate for the size of this hidden population if we assume
that counterparts of the $\sim$20 unidentified field-center sources (expected
to be real) have masses larger than $\sim 1~M_{\odot}$ (cf. Figure
\ref{fig:noid}) and that the ratio of the number of stars with $M >
1.0~M_{\odot}$ to stars with $M < 1.0~M_{\odot}$ is the same as for our {\em
optical sample}, namely, $\sim$25\% for the field center and $\sim$16\% for the
entire FOV.  We obtain the interesting result that the undiscovered population
numbers between 80 and 125 stars, i.e., about half of the optically-well
characterized members \footnote{About 200 stars are placed in the HR diagram,
and $\sim$140 are in our {\em optical sample}.} in the same region and about a
quarter of the optical objects studied by \citet{hil97}.  For comparison, there
are $\sim 600$ stars seen in the infrared with $K < 14.0$, implying masses that
roughly correspond to the mass range spanned by our {\em optical sample}. We
conclude, therefore, that the central region of the ONC contains $\sim$100
additional stars that remain hidden from the deepest $K$-band surveys conducted
to date.

\section{Discussion and Summary \label{sect:disc}}

Using {\em Chandra} X-ray data and recent optical/IR data from the
literature, both presented in Paper~I, we have studied relationships
among X-ray activity indices and various stellar parameters for an
optically-selected sample of ONC members. Our ultimate goal is to
explain on physical grounds the mechanisms responsible for X-ray
emission and magnetic activity in PMS stars.

We do not find any correlation of either $L_X$ or $L_X/L_{bol}$ with stellar
rotational period. This result has sometimes been regarded as evidence against
the $\alpha$ - $\omega$ dynamo explanation of activity for PMS stars. We note
however that: 1) Although the sample of stars for which we have a measure of
rotational period is larger than ever considered in similar past studies, it is
still biased toward more active stars, as judged from distributions of X-ray
activity indicators; 2) Many of the sources with known rotational properties may
thus have saturated activity levels and the intensity of their X-ray emission
may therefore be insensitive to the rotational period; 3) 90\% of these
rotational periods are shorter than 10 days; 4) As discussed by
\citet{fla02}, periods up to $\sim 16$ days might correspond to saturated
activity for $\sim$1~Myr-old stars, according to the {\em consolidated}
activity~--~Rossby~number relation.  We therefore do not regard the present lack
of evidence of a rotation-activity relationship as conclusive for the ONC and look forward the
realization of more complete rotational period databases.  We also note that
even if the measured rotational periods were representative of the whole ONC
population (as recently indicated by \citealt{rho01}), a lack of correlation does not necessarily require a new explanation
of activity.

As previously noted for other star forming regions \citep[e.g.,][]{fei93}, we
find a significant correlation of $L_X$ with mass for $M \lesssim
3.0~M_{\odot}$. The logarithm of the median $L_X$ rises quite linearly on a
logarithmic mass scale from $Log(M/M_{\odot}) \sim -0.9$ to $\sim 0.4$, but for
higher masses, drops sharply before rising again above $Log(M/M_{\odot}) \gtrsim
1.0$. The drop in activity at $M \sim 3.0~M_{\odot}$ is even more dramatic in
$L_X/L_{bol}$ (more than two orders of magnitude) and may be further enhanced if
proper account for multiplicity were taken. Supported by recent PMS evolutionary
models (SDF) we interpret this drop as the effect of the disappearance of a
convective envelope for stars more massive than $\sim 3~M_{\odot}$. In general
we note a convincing correlation between the presence of a convective layer and
activity for $M < 10~M_{\odot}$.


For $0.5 \lesssim M/M_{\odot} \lesssim 3.0$, the median $Log(L_X/L_{bol})$ is
observed to be quite stable at about -3.5, lower than (though close to) the
saturation level. At lower masses however we observe a decrease of the median
$L_X/L_{bol}$ with decreasing mass, as it falls by a factor of $\sim 2.8$
($0.45$ dex) between $M=0.5-1.0$ and $M=0.1~M_{\odot}$. We are unsure of the
origin of this mass dependence but note a rough coincidence between the turning
point in the $L_X/L_{bol}$ - mass relation, at $\sim 1~M_{\odot}$, and the mass
at which 1~Myr stars cease to be fully convective (cf. Figure
\ref{fig:LXLbvsM4}).  We suggest this may be due to a transition from an
$\alpha$--$\omega$ dynamo to a fully-convective dynamo.  It may be even more
relevant to note that the median value of the Ca~II line equivalent width,
which can be used as a proxy of disk accretion (cf. \S \ref{sect:CaII}), is
significantly lower for stars with masses below 0.5$M_\odot$ (median
$EW$=-0.15) than for stars with mass between 0.5 and 3.0$M_\odot$ (median
$EW$=0.94), possibly indicating a larger fraction of accreting stars at low
masses\footnote{There is indeed growing evidence that disk lifetime is longer
for lower mass stars. Two examples of such a result, based on detection of
disks in the IR, can be found in \citet{hil98a} for the ONC and \citet{hai01}
for IC~348.}. We therefore propose (cf.  \citealt{fla02}) that the increase of
$L_X/L_{bol}$ at low masses {\em might} be the effect of a decrease, with
increasing stellar mass, of the fraction of stars that are accreting and/or are
surrounded by disks. We indeed found such stars to have significantly lower
activity levels (\S \ref{sect:CaII}).

Although we do not detect any securely identified brown dwarfs, we are able to
estimate the mean X-ray luminosity for the 9 spectroscopically confirmed BDs of
the BD bin in Figure \ref{fig:BD}. The mean $L_X$ of $\sim
10^{28.5}$~erg$\cdot$s$^{-1}$ appears to be on the same $L_X$ vs. mass
relationship we observe for higher mass stars. This is not particularly
surprising because at this evolutionary stage BDs do not fundamentally differ
from low mass fully convective PMS stars.

We have evidence of the presence of a population of very weak emitters at the
low mass end. In the $0.25-0.50~M_{\odot}$ mass bin, for example, we observe
that the mean X-ray luminosity of stars individually not detected in the X-ray
data (Paper~I) is $\gtrsim 1.5$ orders of magnitude lower with respect to the
mean of the whole sample, indicating either a very wide or a bimodal X-ray
luminosity distribution. At least part of this dispersion can be explained by a
dependence of activity on circumstellar accretion rate. By discriminating stars
with high and low circumstellar accretion on the basis of the Ca~II ($\lambda =
8542\AA$) equivalent width, we prove with high statistical significance that
the X-ray activity of the two groups differs by as much as an order of
magnitude. The relationships of median $L_X$ and $L_X/L_{bol}$ with mass appear
to hold exclusively for the low accretion sample.  We also note that the
scatter in activity at a given mass appears to be larger for the high accretion
sample than for the low accretion sample. Although the origin of these
differences is presently unknown, we put forward three possible scenarios:

1) Accreting stars are rotating slowly due to disk breaking and their dynamo
efficiency depends on rotation, while those accreting less have broken free of
their disks to spin-up and thus saturate their activity.  This hypothesis seems
to be contradicted by two facts: low and high accretion stars have
statistically indistinguishable rotational periods (cf. Figure
\ref{fig:P_dist}) and, as  reported in \S \ref{sect:prot}, we do not observe a
correlation   between activity and $P_{rot}$. However, the sample of stars for
which we have rotational information is not complete and may be subject to
biases (see \S \ref{sect:prot}), so that we cannot rule out this  possibility
until a more representative sample of rotational periods becomes available.

2) Accretion and/or the presence of a disk, and/or  outflows influences
coronal geometry, for example, by decreasing the fraction of the stellar
surface available for the closed magnetic structures from which X-ray emission
is thought to emanate. Other scenarios of altered geometries have been proposed; e.g.,
\citet{mon00} argue that coronal structures might extend to the
inner part of disk. Because of the inhomogeneous and time
variable nature of accretion, variability studies and simultaneous
X-ray/optical observations could help clarify this matter. 

3) Accreting stars have higher X-ray extinction than assumed, so that the
difference in inferred $L_X$ and $L_X/L_{bol}$ is only apparent. Although in
Paper~I our ACIS spectra analysis was not able to exclude this possibility with
high confidence, due mainly to the low photon statistics of  sources associated
with high accretion stars, we also found no supporting evidence for this
scenario. Moreover, an empirical determination of the $N_H$ vs. $A_V$ relation
for the $\rho$ Ophiuchi population (cf. \citealt{ima01}) also does not support
this hypothesis. The reason for the departure from the assumed $A_V$-$N_H$
relation could be that the gas to dust ratio may differ for accreting stars
with respect to the average interstellar value, resulting in an underestimation
of the X-ray absorption. One could imagine a scenario in which  accretion gas
columns that cross the line of sight would obscure X-rays and let optical/IR
radiation through. The most obvious test for this hypothesis is a better
determination of hydrogen absorbing columns through deeper (i.e., longer
exposures) medium spectral resolution X-ray observations. As for the previous
hypothesis, variability studies might also give useful clues.

Finally we have investigated the nature of sources detected in our HRC X-ray
data and not identified with any optical/IR object. On the basis of their X-ray
spectra and location in the cloud we conclude that, as a class, they are subject
to high extinction. We exclude an extragalactic nature for most of these objects by
considering the sky density of such objects as a function of X-ray flux and of
the molecular cloud total extinction and conclude that most unidentified X-ray
sources are likely associated with deeply embedded, high intrinsic X-ray
luminosity, ONC members. Given the relationship between $L_X$ and stellar mass
we infer that these stars are most likely fairly massive ($M \gtrsim
1.0~M_\odot$), thus indicating the likely presence of a significant number of
lower mass, as yet undetected ONC members.


In light of these results, we propose this scenario:

1) Activity in $\sim 1~Myr$ old low mass PMS stars is ultimately due to a dynamo
mechanism that requires a convective layer in order to function.
Rotation may or may not be a fundamental ingredient of this dynamo. 

2) The same mechanism responsible for X-ray emission of low mass stars is
probably also at work in brown dwarfs.

3) Many of our stars appear to be saturated. The fraction of saturated stars,
or alternatively the saturation level, depends on stellar mass. At low masses
we might be seeing the effect of decreasing accretion/disk fraction with
increasing mass, or the transition between two kinds of dynamos:  one that
functions on fully convective stars and brown dwarfs and in which the
saturation level depends on mass, and another at work on partially convective -
partially radiative stars, characterized by a constant saturation level.

4) Highly accreting  stars are seen to be less active with respect to low
accretion counterparts, possibly because of (a) different rotational
properties, (b) disk/accretion induced modifications of the coronal magnetic
field geometry, or (c) anomalous interstellar X-ray absorption due to the
presence of circumstellar material.

Additional deep X-ray and IR observations are needed to advance our
understanding and enable us to identify which scenario(s) Nature prefers.

\section*{Acknowledgments}

The authors would like to thank L. Hartman for useful discussion.

This work was partially supported at the CfA by NASA contracts
NAS8-38248 and NAS8-39073 and by NASA grant NAS5-4967. F.D., E.F.,
G.M., and S.S. wish to acknowledge support from the Italian Space
Agency (ASI) and MURST. E.F. would like to thank the CfA for its
hospitality during his Fellow visits.

\appendix
\section{``Composite'' source analysis \label{app:nodet}}

Here we describe a ``composite'' source method we employed to study the mean
X-ray properties of groups of ONC members, some or all of which were
individually undetected in our HRC data. The method was applied in \S
\ref{sect:LxvsMass} to undetected low mass stars, and in \S \ref{sect:bd}, to
objects of substellar mass.

The basic steps of our method are to: 1) define a suitable sample of stellar or
substellar objects; 2) determine the number of photons detected in a circular
region at the X-ray position corresponding to each {\em optical} object; 3)
correct the photon numbers for background contributions and the fact that a
(predictable) fraction of the photons from a point-like source will fall outside
the extraction circle; 4) sum the contributions of single objects and derive a
signal-to-noise ratio (SNR) for the ``composite'' source; 5) divide the
``composite'' source net counts (or upper limit, for SNRs below a set threshold)
by the cumulative exposure time to obtain a mean count rate for objects in the
sample; and finally, 6) convert the mean count rate (or upper limit) to a
``composite'' $L_X$ value.

The critical points in these six steps for estimating $L_X$ values for the
composite sources are as follows:

\paragraph{1)} The selection of composite source samples was based on
stellar (or substellar) mass as given by our {\em optical sample}. We also added
three more requirements: a) that the objects lay in the inner 5$^{\prime}$ of
the HRC field of view, b) that the closest detected X-ray source (other than the
object in question, if detected) was farther away than 5$^{\prime\prime}$ and c)
that the measured optical extinction $A_V$ was less than 2.0.  Constraint (a)
ensures that the PSF at the position of all the objects in the sample contains
more than $\sim 60\%$ of the photons falling within a circle of
$2^{\prime\prime}$ radius, while (b) isolates the source to avoid
contamination with photons from neighboring sources.  Constraint (c) reduces the
uncertainty in the conversion between composite source count rates and X-ray
luminosities.

\begin{figure}
\centerline{\psfig{figure=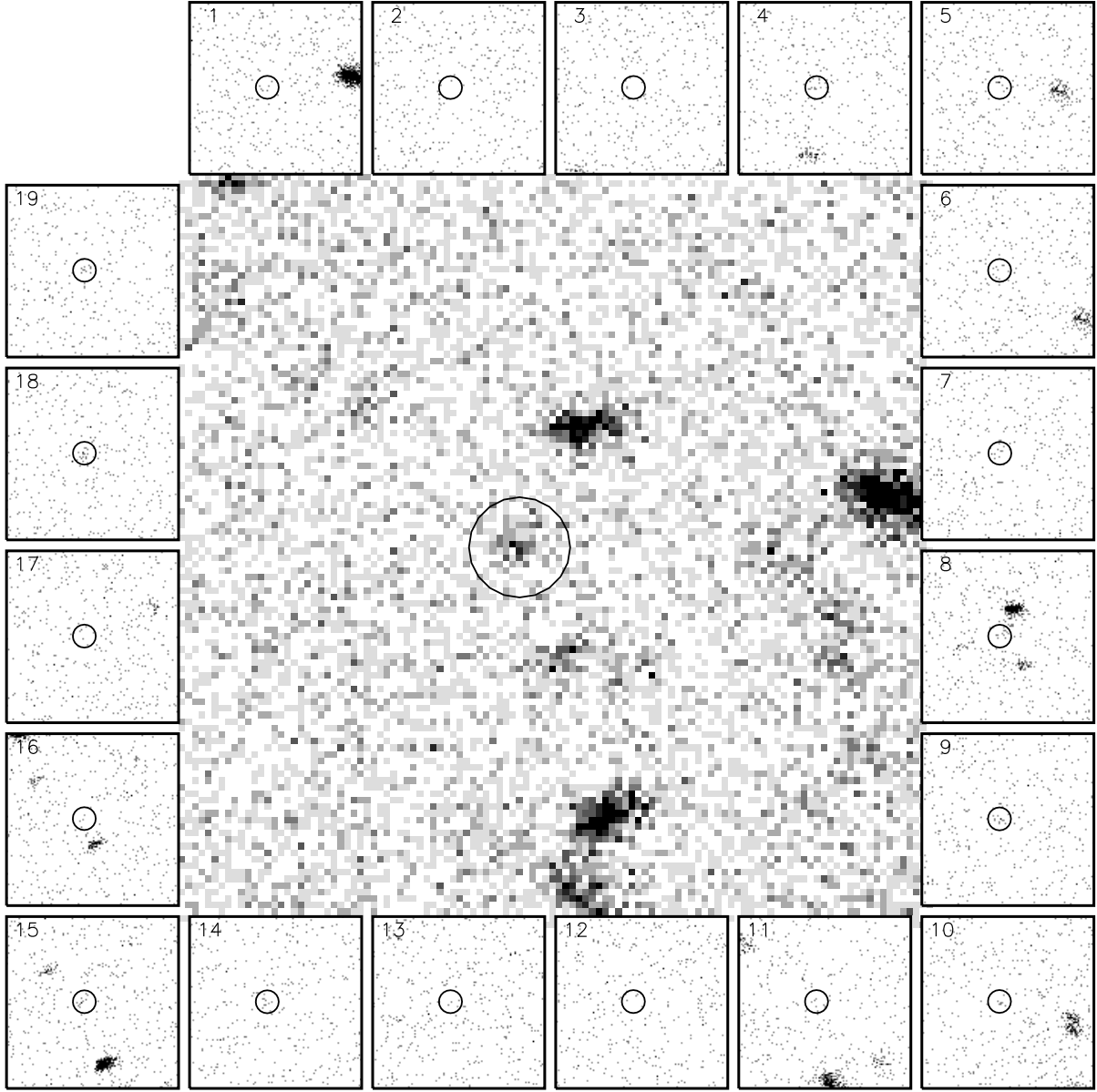,width=16cm}}
\caption{Center: Composite image of 19 undetected stars with masses in
the $0.16-0.25~M/M_{\odot}$ range (see \S \ref{sect:bd}). Edges: HRC
images at positions of 19 stars.  Center and edge finding charts all
have the same angular extent of $30^{\prime\prime}\times30^{\prime\prime}$.
Circles indicating optical positions all have the same radius of
$2^{\prime\prime}$. \label{fig:BD_im}}
\end{figure}

\paragraph{2)} Individual source photons were extracted from a
2$^{\prime\prime}$ circle centered on the optical star position.  Figure
\ref{fig:BD_im} shows an example of the extraction regions for undetected
members in the $0.16-0.25~M/M_{\odot}$ mass range as well as a graphical
representation of the composite source resulting from the summation process.

\paragraph{3)} The background contribution in each of the extraction circles
was estimated from the values of the source-free background map produced by
{\sc pwdetect} (cf. Paper~I). The fraction of counts from the object falling
within this extraction circle ranges from $\sim$ 60\% to 95\% and was estimated
from the images of calibrated PSFs obtained for the position of each source
using the {\sc ciao} task {\sc mkpsf} and the F1 standard PSF library file
\citep{kar01}. Sources were assumed for this purpose to be monochromatic with
an energy of 2.0 keV. A visual comparison of the expected and observed shape of
the PSF for bright sources in our present sample indicates a good qualitative
match.

\paragraph{4)} From counting statistics we estimated the probability that the
measured total number of extracted counts is the result of random fluctuation
in the background level. In those cases in which such probability was larger
than 0.135\% ($3\sigma$), we computed upper limits to the composite source
counts using the same confidence threshold. 

\paragraph{5)}The cumulative exposure time used to compute mean count rates for
the objects in each mass segregated sample was obtained by summing the values
of the exposure map (cf. Paper~I) at the position of the objects. 

\paragraph{6)} We converted mean count rates to X-ray luminosities in a manner
consistent with Paper~I, but such conversion depends critically on extinction.
While we can derive conversion factors for each of the component objects, a
global conversion factor for a composite source cannot be readily defined. If
all the objects in the sample had the same intrinsic X-ray luminosity the
conversion factor for the average source would be the average of the single
conversion factors. In order to reduce the range of individual conversion
factors we have limited our object sample to low optical extinction ($A_V <
2.0$). However some uncertainty remains: as shown in Figure 8 of Paper~I, this
range of $A_V$ corresponds to a factor of about 3 ($\sim 0.5$ dex) variation in
the conversion factor at our chosen source temperature (2.16~keV). We therefore
report two values of luminosity for each of our average sources: one computed
with the mean of the conversion factors, the other with the median.

The results of applying this technique are discussed in \S \ref{sect:LxvsMass}
and \S \ref{sect:bd} and are shown in Figure \ref{fig:BD}.

\end{document}